\documentclass{article}

\usepackage{arxiv}

\usepackage{amsmath}
\usepackage{amssymb,times}
\usepackage{subfigure}
\usepackage{color}
\usepackage{bm}

\usepackage[utf8]{inputenc} 
\usepackage[T1]{fontenc}    
\usepackage{hyperref}       
\usepackage{url}            
\usepackage{booktabs}       
\usepackage{amsfonts}       
\usepackage{nicefrac}       
\usepackage{microtype}      
\usepackage{cleveref}       
\usepackage{lipsum}         
\usepackage{graphicx}
\usepackage{natbib}
\usepackage{doi}

\title{A Spectral-Spatial Fusion Anomaly Detection Method for Hyperspectral Imagery}

\author{ \href{https://orcid.org/0000-0001-6181-2326}{\includegraphics[scale=0.06]{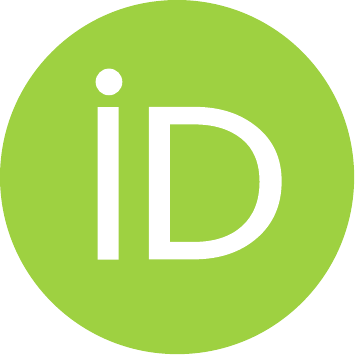}\hspace{1mm}Zengfu~Hou}\thanks{website: https://zephyrhours.github.io} \\
    School of Information and Electronics\\
	Beijing Institute of Technology\\
	Beijing, 100081\\
	\texttt{zephyrhou@126.com} \\
	\And
	{\hspace{1mm}Siyuan~Cheng} \\
	Space star technology co., LTD\\
	Beijing, 100086\\
	\texttt{347707597@qq.com} \\
	\And
	{\hspace{1mm}Ting~Hu}\thanks{Corresponding author} \\
	School of Information and Electronics\\
	Beijing Institute of Technology\\
	Beijing, 100081\\
	\texttt{ht\_hc2@163.com} \\
}

\renewcommand{\shorttitle}{A spectral-spatial fusion anomaly detection method for hyperspectral imagery}

\hypersetup{
pdftitle={A Spectral-Spatial Fusion Anomaly Detection Method for Hyperspectral Imagery},
pdfsubject={q-bio.NC, q-bio.QM},
pdfauthor={Zengfu Hou, Siyuan Cheng, Ting Hu},
pdfkeywords={Hyperspectral, Anomaly detection, Saliency weight, Feature enhancement, Adaptive fusion}
}

\begin{document}
\maketitle
\begin{abstract}
In hyperspectral, high-quality spectral signals convey subtle spectral differences to distinguish similar materials, thereby providing unique advantage for anomaly detection. Hence fine spectra of anomalous pixels can be effectively screened out from heterogeneous background pixels. Since the same materials have similar characteristics in spatial and spectral dimension, detection performance can be significantly enhanced by jointing spatial and spectral information. In this paper, a spectral-spatial fusion anomaly detection (SSFAD) method is proposed for hyperspectral imagery. First, original spectral signals are mapped to a local linear background space composed of median and mean with high confidence, where saliency weight and feature enhancement strategies are implemented to obtain an initial detection map in spectral domain. Futhermore, to make full use of similarity information of local background around testing pixel, a new detector is designed to extract the local similarity spatial features of patch images in spatial domain. Finally, anomalies are detected by adaptively combining the spectral and spatial detection maps. The experimental results demonstrate that our proposed method has superior detection performance than traditional methods.
\end{abstract}


\section{Introduction}
Hyperspectral images (HSI) are different from multispectral or panchromatic images, which record not only spatial distribution information of objects but also its spectral structure information. Hence, different materials can be distinguished by using their spectral features \cite{hou2018novel,hou2022joint}. These distinctive fine spectra like human fingerprints with uniqueness have shown great potential in target detection applications \cite{zhao2021hyperspectral}. 

However, prior knowledge of target spectrum is difficult to obtain in advance, which limits its application scenarios. In practical terms, by only analyzing spectral signals of hyperspectral images, it is possible to identify different objects. Therefore, hyperspectral imaging has shown great potential applications for various scenarios, such as scene classification \cite{hu2017fusionet}, spectral unmixing \cite{villa2010spectral}, change detection \cite{hou2021hyperspectral,hou2021three}, target detection \cite{zhao2021hyperspectral}, and anomaly detection \cite{liu2021multipixel,hou2020background}, etc.. Among these applications, anomaly detection has gotten a lot of attention due to its ability to recognize objects differing from the sourrounding background only using spectral differences rather than prior spectrum \cite{houcollaborative}. 

Hyperspectral anomaly dection aims to identify objects of interest that are spatially or spectrally different from its surrounding. Therefore, how to effectively select and mark anomalous pixels from an entire image has become the main challenge. For this challenge, various statistical and physical models have been reported.

Assuming that background obeys multivariate Gaussian normal distribution, well-known Reed-Yu (RX) \cite{reed1990adaptive} algorithm is proposed, where Mahalanobis distance is calculated to measure the similarity between testing pixel and background ones. When the whole image is considered as Gaussian background model, it is called global RX (GRX) \cite{reed1990adaptive}. If RX detector estimates background model using local statistics, it is called local RX (LRX) \cite{molero2013analysis}. However, the background cannot be simply described with a multivariate normal distribution because of its extremely complicated distribution. Therefore, some improved RX-based methods such as weighted RX (WRX) \cite{guo2014weighted}, linear filter RX (LFRX) \cite{guo2014weighted}, and kernel RX (KRX) \cite{kwon2005kernel} are investigated. 

Compared with the above background statistics-based methods, some linear representation-based methods producing better detection performance have been proposed. For example, the collaborative representation detector (CRD) \cite{li2014collaborative} exploiting the similarity relationship between non-anomalous pixel and background pixels had achieved satisfactory detection performance, which represented the testing pixel linearly via its neighborhood pixels. However, the outliers in background pixels could degrade the representation accuracy. To generate relatively pure background pixels, a CRD with outlier removal anomaly detector (CRDBORAD) \cite{vafadar2017hyperspectral} was proposed, removing some suspected anomaly pixels. Subsequently, a CRD combined with principal component analysis (PCAroCRD) \cite{su2018hyperspectral} was put forward, where principal component analysis (PCA) was used for outlier removal. Considering the contributions of spatial information, a CRD-based local summation and inverse distance weight detector (LSADCRIDW) \cite{tan2019anomalysubspace} was proposed to makes full use of the various local spatial distribution information of the neighboring pixels. Utilizing the least squares technique to purify background pixels as well as weighting the average saliency of the neighboring pixels into the detection map, CRD with background purification and saliency weight [10] (CRDBPSW) was put forward. 

The third anomaly detection method, as the low-rank matrix decomposition serves as a powerful tool to exploit the intrinsic low-rank property of images \cite{candes2011robust}. These low rank representation (LRR)-based methods aim to find a low-dimensional subspace to represent the entire image with as few as possible atoms. Assume that the background pixel obeied a single subspace, the classical robust principal component analysis (RPCA) \cite{candes2011robust} method that decomposed the image matrix into a low-rank matrix and a sparse matrix, was proposed,  where the sparse matrix was used for anomaly detection \cite{liu2012robust}. By applying the RX detector to the sparse matrix, the RPCA-RX \cite{liu2012robust} detector was proposed to perform anomaly detection. Different from RPCA, the LRR model used multiple subspaces to represent the data, with being more suitable for an HSI with complex background features.  Subsequently, by adding a sum-to-one constraint,  the low-rank representation sum-to-one (LRRSTO) \cite{xu2015novel} mode was proposed for anomaly detection. Considering that the direct application of LRR model was sensitive to a tradeoff parameter, the LRR-based and Learned Dictionary (LRRaLD) \cite{niu2016hyperspectral} method was proposed, where the dictionary was learned from the entire image with a random selection processing. To make full use of the local background statistics information, single/multiple local windows-based LRRSTO (SLW/MLW\_LRRSTO) \cite{tan2019anomalylowrank} method was proposed. As yet, many LRR-based methods, such as sparse representation (SRD) \cite{chen2011simultaneous}, low-rank and sparse representation (LRASR) \cite{xu2015anomaly}, and graph total variation regularized low rank representation (GTVLRR) \cite{cheng2019graph}, etc. have been published with better detection results.

Recently, deep learning-based methods have drawn increasing attention in hyperspectral anomaly detection. A transferred deep convolutional neural network detector (CNND) \cite{liwei2017cnnd} was the first attempt to train an anomaly detector using pixel pairs selected from a reference data. With the birth of autoencoders (AEs), various AEs such as spectral constrained adversarial AE (SC\_AAE) \cite{xie2019spectral}, stacked denoising AEs (SDAs) \cite{zhao2018spectral}, and manifold-constrained AE network \cite{lu2019exploiting}, etc. were introduced into hyperspectral anomaly detetion \cite{wangshaoyu2022AutoAD,zhao2017hyperspectral}. They usually achieved outstanding judge for anomalous pixels via the spectral signal; nevertheless, the spatial features were ignored. 

To address the above problems, an effective spectral-spatial fusion anomaly detection (SSFAD) framework is designed in this papar. In the spectral domain, a local median-mean line (LMML) projection method is proposed, where local adjacent pixels surrounding the testing pixel are mapped into the median-mean line to rectify the position of the background samples. In addition, a monotonic increasing function based on inverse distance weight (IDW) \cite{tan2019anomalysubspace} is used to further increase the gap between the anomalous and the local background pixels. The spectral feature map is obtained by a saliency weight-based detector. In the spatial domain, a local spatial similarity measurement method is proposed to fully use of spatial similarity features of local background. Finally, an adaptive score strategy is proposed to determine the fusion coefficients of the two complementary parts.

The main contributions can be summarized as follows.

1) A novel dual-pipeline framework is proposed for hyperspectral anomaly detection, where spectral and spatial
features are fused by the proposed adaptive score strategy to further highlight anomaly.

2) In spectral domain, LMML projection and feature enhancement strategies were proposed to increase the discrimination between anomalious and background pixels.

3) When extracting spatial features, a local spatial similarity measurement method is designed to improve the detection effect by exploiting intrinsic local pixels relation.

The remainder of this article is organized as follows. In Section II, a detailed description of the proposed framework is presented. In Section III, four real datasets are utilized to verify the proposed method, and the parameters and results are analyzed and discussed. The conclusion is drawn in Section IV.

\section{Proposed Detection Method}
\begin{figure*}
	\centering
	\includegraphics[width = 5 in]{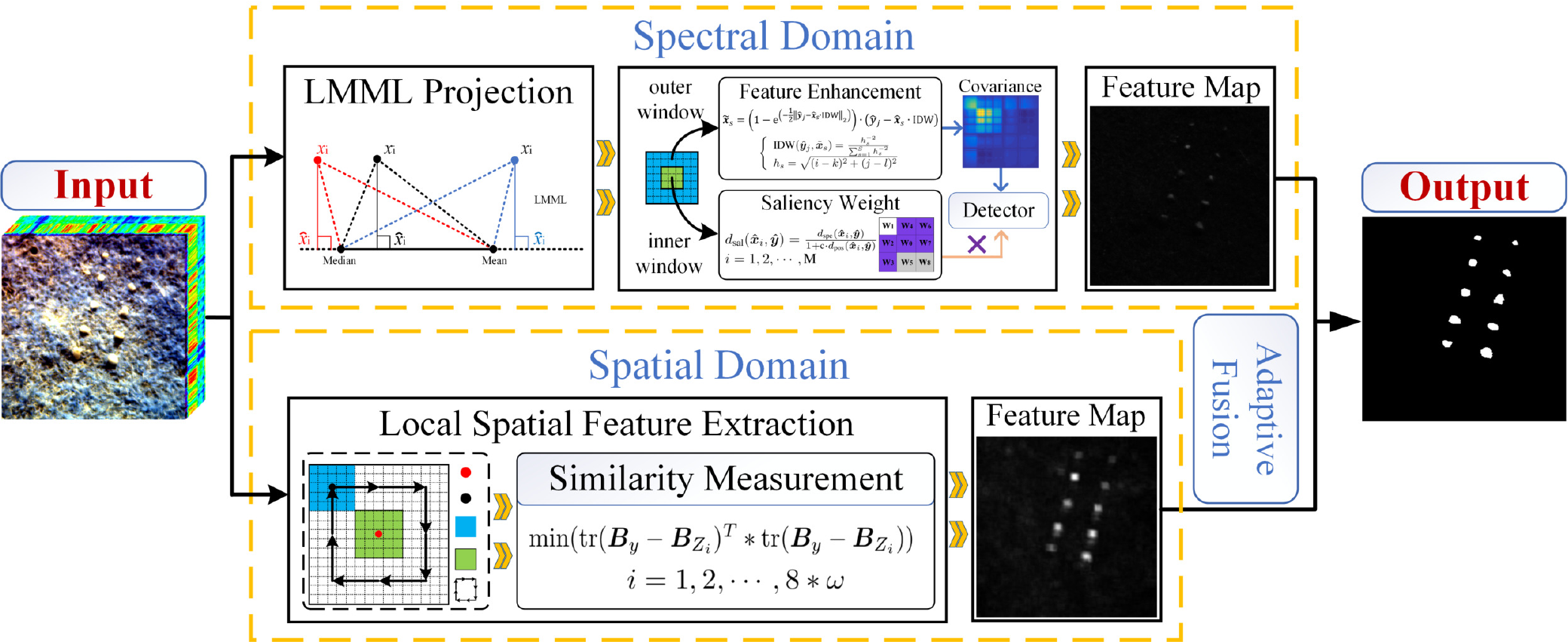}
	\caption{Framework of proposed SSFAD detector for hyperspectral anomaly detection. 
		\label{Fig:Workflow}}
\end{figure*}

During background modeling in traditional methods, anomalous pixels are judged by the spectral signals, while the spatial features were ignored. Homogeneous pixels in the background may affect the detection performance. Thus, it is necessary to maximize the spatial texture between background and anomaly. Fig. \ref{Fig:Workflow} displays a flowchart of the proposed SSFAD framework, which consists of the following steps. To begin, spectral feature map is obatined by LMML, feature enhancement and saliency weight strategies. After that, spatial feature map is extracted by local spatial similarity measurement. Finally, two complementary detection maps are adaptive fused to highlight anomaly.

\subsection{Spectral Domain-Based Anomaly Feature Extraction}
\begin{figure}
	\centering
	\includegraphics[width= 3 in]{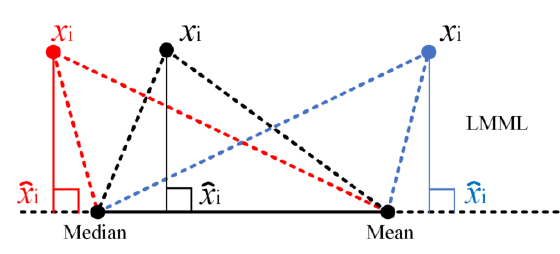}
	\caption{Illustration of the projection of testing pixel $\bm{x}_i$ on LMML.}
	\label{Fig:LMML}
\end{figure}

In anomaly detection, covariance statistic-based methods rely excessively on the accuracy of background statistics. In hyperspectral imaging, the spectral curve is easily disturbed by various complex factors such as noise and dark current causing the oscillation of spectral curves. Moreover, in real scenarios, anomalous targets play outlier roles to contaminate the background data, which hinders the accurate estimation of the background pixels. Both phenomena could seriously degrade the performance of covariance statistic-based anomaly detection methods. In this paper, to degrade the negative effect on the mean and covariance matrix caused by anomalious pixels \cite{imani2017rx}, a LMML projection method is proposed to rectify the position of background samples. Specifically, all the neighborhood pixels inside the outer window $\omega_{\text{out}}$ of testing pixel are projected into LMML space. 
\begin{equation}
	\begin{array}{l}
		$$  \bm{\hat{x}}_i = (1-\eta_i) \bm{M} +\eta_i\bm{m}, \ i=1,2,\cdots, N   $$, 
	\end{array}
\end{equation}
where $\bm{M}$ and $\bm{m}$ are the median value and mean value, respectively. $\eta_i \in $ [0, 1] denotes the position parameter. $N = \omega_{\text{out}}*\omega_{\text{out}}$, which represents the number of pixels in the local window centered on the testing pixel $\bm{x}_i$. As shown in Fig. \ref{Fig:LMML}, the connecting line between pixel $\bm{x}_i$ and its projection point $\bm{\hat{x}}_i$ is perpendicular to the median-mean line. According to the geometric theory, there is,
\begin{equation}
	\begin{array}{l}
		$$  [(1-\eta_i) \bm{M} +\eta_i\bm{m}-\bm{x}_i]\cdot(\bm{m}-\bm{M})=0  $$, 
	\end{array}
\end{equation}
then,
\begin{equation}
	\begin{array}{l}
		$$  \eta_i = \frac{(\bm{x}_i-\bm{M})\cdot(\bm{m}-\bm{M})}{(\bm{m}-\bm{M})\cdot(\bm{m}-\bm{M})} , \ i=1,2,\cdots, N   $$. 
	\end{array}
\end{equation}

After the LMML operation, local pixels are mapped in the median-mean direction, while outliers are distributed in sides of the median-mean line. The farther away from the median and mean points, the higher the probability of anomalies. Furthermore, an adaptive lcoal background prototype is obtained via the LMML projection, where the projection point of each pixel is used as the rectified background pixels to calculate the local covariance matrix.

Inspired by \cite{lei2019spectral}, where a suppression function is used to construct a discriminative feature space. Hence, to further enlarge the difference between background and anomalous pixels, a novel monotonic increasing function is proposed, which is formulated as,
\begin{equation}
	\left\{
	\begin{array}{l}
		$$  \tilde{\bm{x}}_s = (1-\exp(-\frac{1}{2}\Vert \bm{\hat{y}}_j -\bm{\hat{x}}_s \cdot \text{IDW}\Vert_2))\cdot(\bm{\hat{y}}_j-\bm{\hat{x}}_s\cdot\text{IDW}) \\
		s = 1,2,\cdots, S $$  
	\end{array},
	\right.
\end{equation}
where $\bm{\hat{y}}_j$ and $\bm{\hat{x}}_s$ represent the projection vectors of the testing pixel and its neighborhood pixels between the outer window $\omega_{\text{out}}$ and the inner window $\omega_{\text{in}}$ in LMML space, respectively. $S=\omega_{\text{out}}*\omega_{\text{out}}-\omega_{\text{in}}*\omega_{\text{in}}$, which is the number of neighborhood pixels. $\text{IDW}$ denotes the inverse distance weight \cite{tan2019anomalysubspace}, which is used to make the most of the space-varying information where the closer the testing pixel is, the higher the similarity between the background pixels and the testing pixel will be. It is formulated as,
\begin{equation}
	\left\{
	\begin{array}{l}
		$$  \text{IDW}(\bm{\hat{y}}_j, \tilde{\bm{x}}_s) = \frac{h^{-2}_s}{\sum_{s=1}^{S}h^{-2}_s}\\
		h_s = \sqrt{(i-k)^2+(j-l)^2} $$  
	\end{array},
	\right.
\end{equation}
where $(i,j)$ and $(k,l)$ denote the geometric coordinates of testing pixel $\bm{\hat{y}}_j$ and its any neighborhood pixel $\tilde{\bm{x}}_s$, respectively. The feature map in spectral domain is obtained by,
\begin{equation}
	\begin{array}{l}
		$$ r = \bm{\hat{y}}\sum^{-1}\bm{\hat{y}}^T$$	
	\end{array}
\end{equation}
where $\sum$ denotes the covariance matrix of the local background pixels $\tilde{\bm{x}}_s$ between the inner window and the outer window.

In addition, considering the influence of adjacent pixels contained in an inner window on the testing pixel, a saliency weight \cite{houcollaborative} is used,
\begin{equation}
	\left\{
	\begin{array}{l}
		$$  d_{\text{sal}}(\bm{\hat{x}}_i,\bm{\hat{y}})= \frac{d_{\text{spe}}(\bm{\hat{x}}_i,\bm{\hat{y}})}{1+\text{c}\cdot d_{\text{pos}}(\bm{\hat{x}}_i,\bm{\hat{y}})}\\
		i = 1,2,\cdots,\text{M}$$  
	\end{array},
	\right.
\end{equation}
where $d_{\text{sal}}$ denotes the saliency distance between the testing pixel $\bm{\hat{y}}$ and its any neighborhood pixel $\bm{\hat{x}}_i$ contained in the inner window. $\text{M}=\omega_{\text{in}}*\omega_{\text{in}} -1 $ is the number of adjacent pixels contained in the inner window. $\text{c}$ is a constant that could cause little effect on the result. As suggested by \cite{houcollaborative}, its value is set to 1. $d_{\text{spe}}$ and $d_{\text{pos}}$ represent the spectral distance and position distance respectively, which are formulated as,
\begin{equation}
	\begin{array}{l}
		$$  d_{\text{spe}}(\bm{\hat{x}}_i,\bm{\hat{y}})= \arccos(\frac{\bm{\hat{x}}_i\bm{\hat{y}}}{\Vert \bm{\hat{x}}_i \Vert_2 \cdot \Vert \bm{\hat{y}} \Vert_2 }) $$
	\end{array},
\end{equation}
\begin{equation}
	\begin{array}{l}
		$$  d_{\text{pos}}(\bm{\hat{x}}_i,\bm{\hat{y}})= \sqrt{(i-k)^2+(j-l)^2} $$
	\end{array}.
\end{equation}

Finally, average all saliency distances to produce the saliency weight of the testing pixel,
\begin{equation}
	\begin{array}{l}
		$$  \bm{w}_{\text{sal}} =\frac{\sum_{i=1}^{\text{M}}d_{\text{sal}}}{\text{M}}  $$. 
	\end{array}
\end{equation}
When the spectral signal of testing pixel is significantly different from that of its adjacent pixels, the pixel is regional saliency. Therefore, the final detection result in spectral domain is reformulated as,
\begin{equation}
	\begin{array}{l}
		$$ r_1 = \bm{\hat{y}}\sum^{-1}\bm{\hat{y}}^T \cdot \bm{w}_{\text{sal}}$$.	
	\end{array}
\end{equation}

\subsection{Spatial Domain-Based Anomaly Feature Extraction}
\begin{figure}
	\centering
	\includegraphics[width= 3 in]{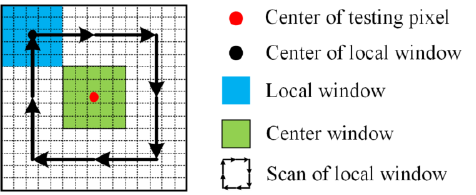}
	\caption{Illustration of calculating pixel's local spatial features.}
	\label{Fig:spatial_features}
\end{figure}

If the testing pixel is anomalous, its saptial structure would differ from ones of the neighborhood background \cite{ju2018hyperspetral}. To calculate the spatial structure similarity between the testing pixel and its neighborhood background, a novel local spatial similarity measurement method is formulated as,
\begin{equation}
	\left\{
	\begin{array}{l}
		$$r_2 =  \min(\text{tr}(\bm{B}_{y}-\bm{B}_{Z_i})^T * \text{tr}(\bm{B}_{y}-\bm{B}_{Z_i})) \\
		i = 1,2,\cdots, 8*\omega $$  
	\end{array},
	\right.\label{eqn:spatial similarity}
\end{equation}
where $\bm{B}_y$ denotes the center window (see the green window in Fig. \ref{Fig:spatial_features}) centered on testing pixel, $\bm{B}_{Z_i}$ is the neighborhood background window (highlighted as the blue window in Fig. \ref{Fig:spatial_features}), and $\omega$ represents the size of spatial structure window.

The local spatial similarity of the center window is calculated following Fig. \ref{Fig:spatial_features}. By sliding the blue window clockwise around the green window, the local spatial similarity information around the center window is obtained. Theoretically, different sizes and shapes should be selected for the spatial structure window for a better adaption to the local characteristics in images. For simplicity, the inner green window is regarded as the spatial structure window. In addition, the size of local background window is set as $\omega$. The most similar spatial structure window are searched from its background area. By Eq. \ref{eqn:spatial similarity}, the spatial structure similarity between center window around testing pixel and its neighborhood background windows is obtained. The bigger the similarity, the smaller the value, and the more likely it is to be background, and vice versa.

\subsection{Spectral-Spatial Fusion Anomaly Detection}
To highlight anomalous targets, two complementary detection maps are adaptively fused. Single feature cannot detect anomaly well in complex scenario, therefore, a fusion strategy is adopted here. A linear score function shown in Eq.\ref{eq13} is employed as the fusion strategy.
\begin{equation}\label{eq13}
	\begin{array}{*{20}{l}}
		$$ R = a*R_1+b*R_2 $$,
	\end{array}
\end{equation}
where $R$ is the finally anomaly detection result. $a$ and $b$ are the corresponding weight factors. It is difficult to tune $a$ and $b$ experiencelly to obtain an optimal value, and a general option is the average pooling operation. That is, $a=b=0.5$. In this paper, a score mechanism is proposed to adaptively determine the values of $a$ and $b$, which is formulated as,
\begin{equation}
	\left\{
	\begin{array}{l}
		$$ a = \frac{\sqrt{\lambda_{\rm max}(R^T_1R_1)}}{\sqrt{\lambda_{\rm max}(R^T_1R_1)}+\sqrt{\lambda_{\rm max}(R^T_2R_2)}} \\
		b = \frac{\sqrt{\lambda_{\rm max}(R^T_2R_2)}}{\sqrt{\lambda_{\rm max}(R^T_1R_1)}+\sqrt{\lambda_{\rm max}(R^T_2R_2)}} $$  
	\end{array},
	\right.\label{eqn14}
\end{equation}
where $\lambda_{\rm max}(\cdot)$ takes the maximal eigenvalue of an matrix. Such adaptive score strategy could help to increase the detection performance.

\begin{figure*}[tp]
	\centering
	\subfigure[\scriptsize{}]{
		\label{Fig:subfig:Gulfport}
		\includegraphics[width=1.5 in]{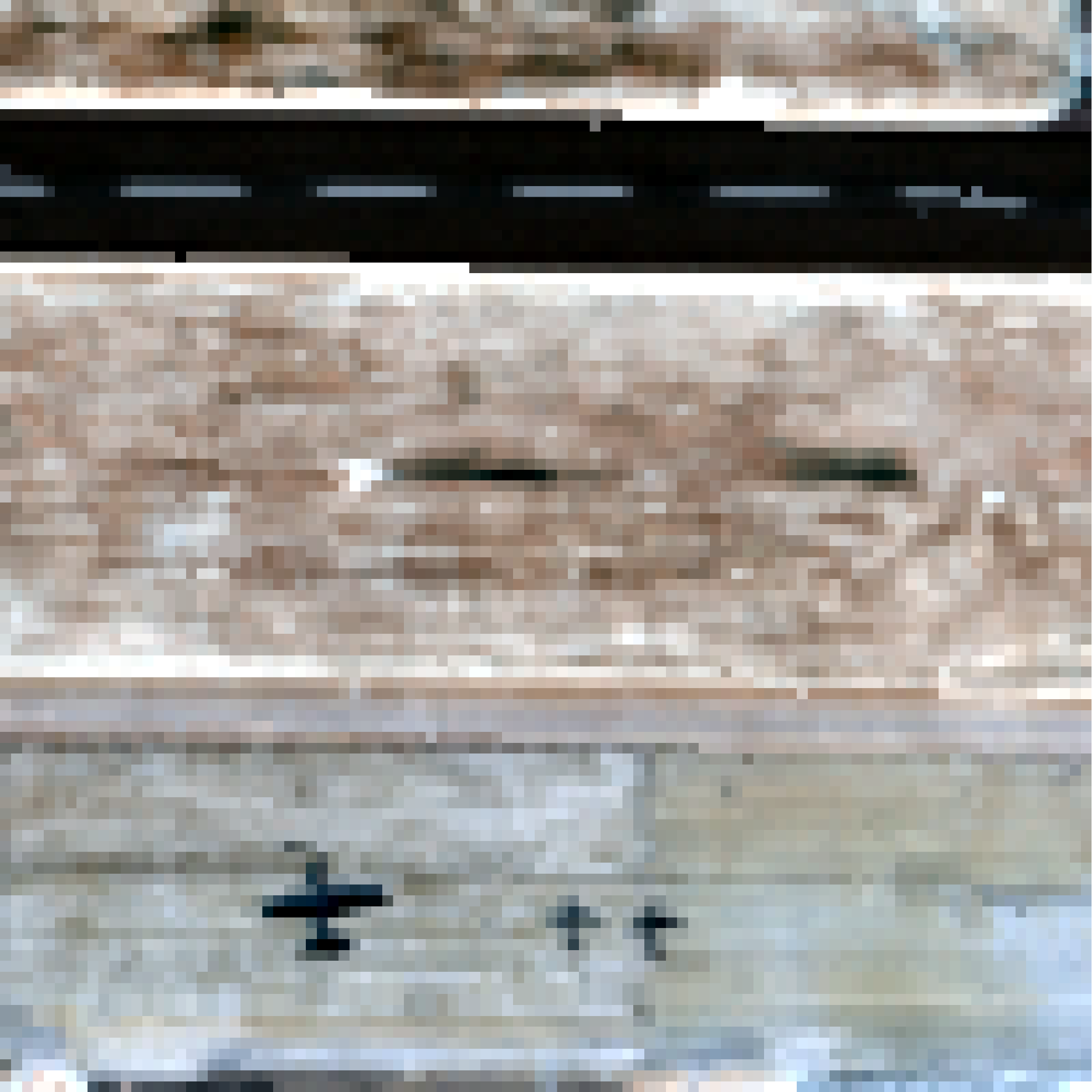}}
	\subfigure[\scriptsize{}]{
		\label{Fig:subfig:PaviaCentra}
		\includegraphics[width=1.5 in]{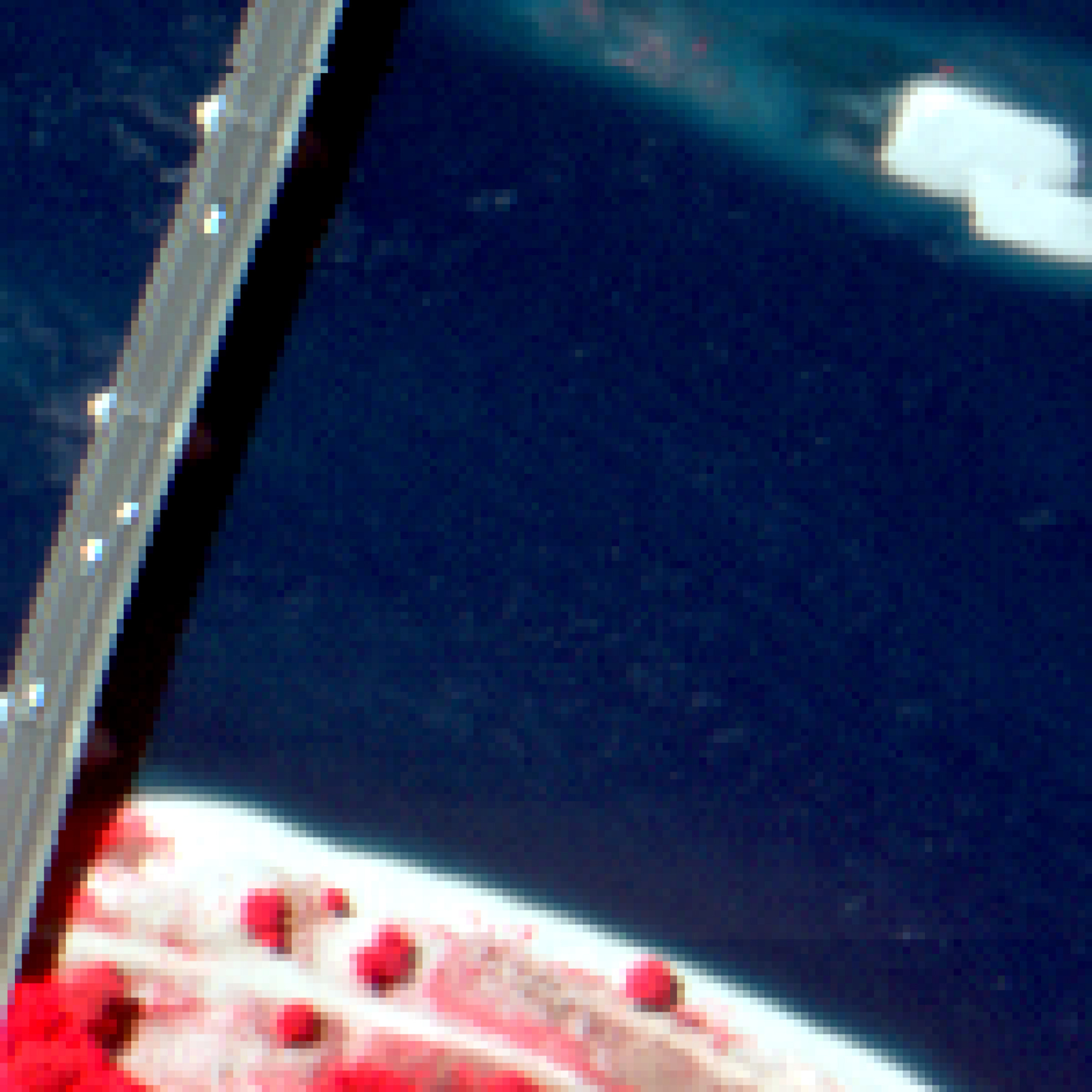}}
	\subfigure[\scriptsize{}]{
		\label{Fig:subfig:Gainesville}
		\includegraphics[width=1.5 in]{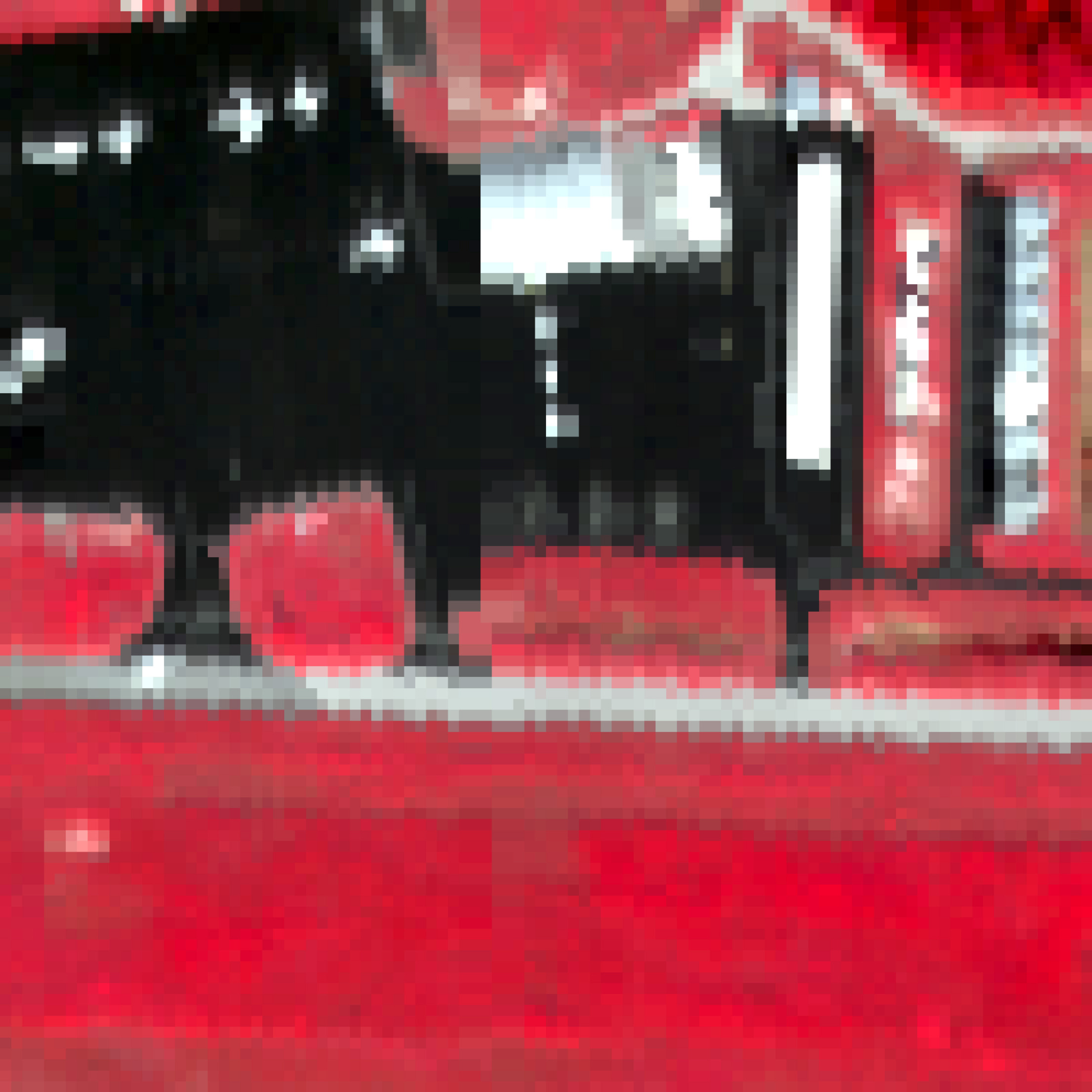}}
	\subfigure[\scriptsize{}]{
		\label{Fig:subfig:Cri}
		\includegraphics[width=1.5 in]{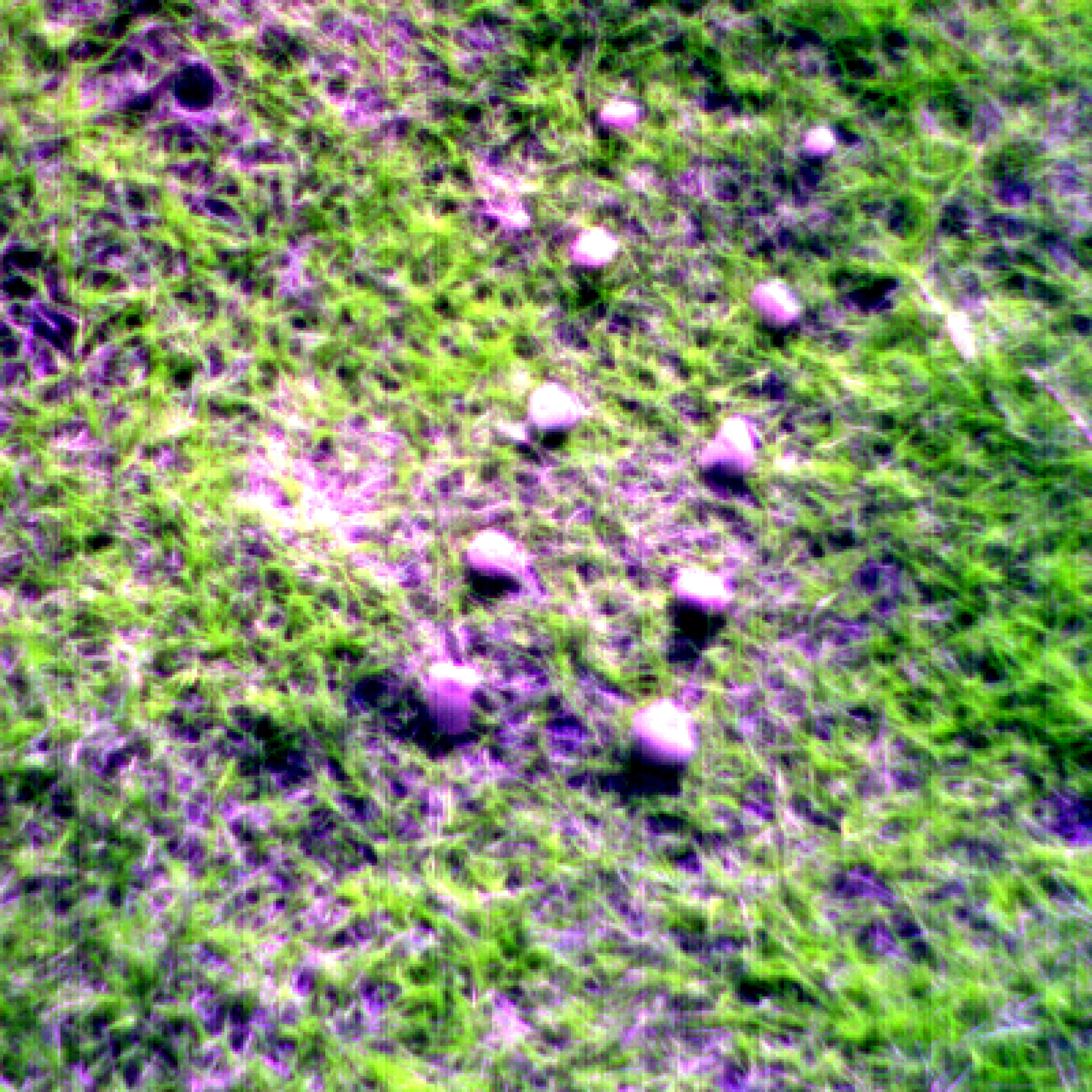}}
	
	\subfigure[\scriptsize{}]{
		\label{Fig:subfig:Gulfport-gt}
		\includegraphics[width=1.5 in]{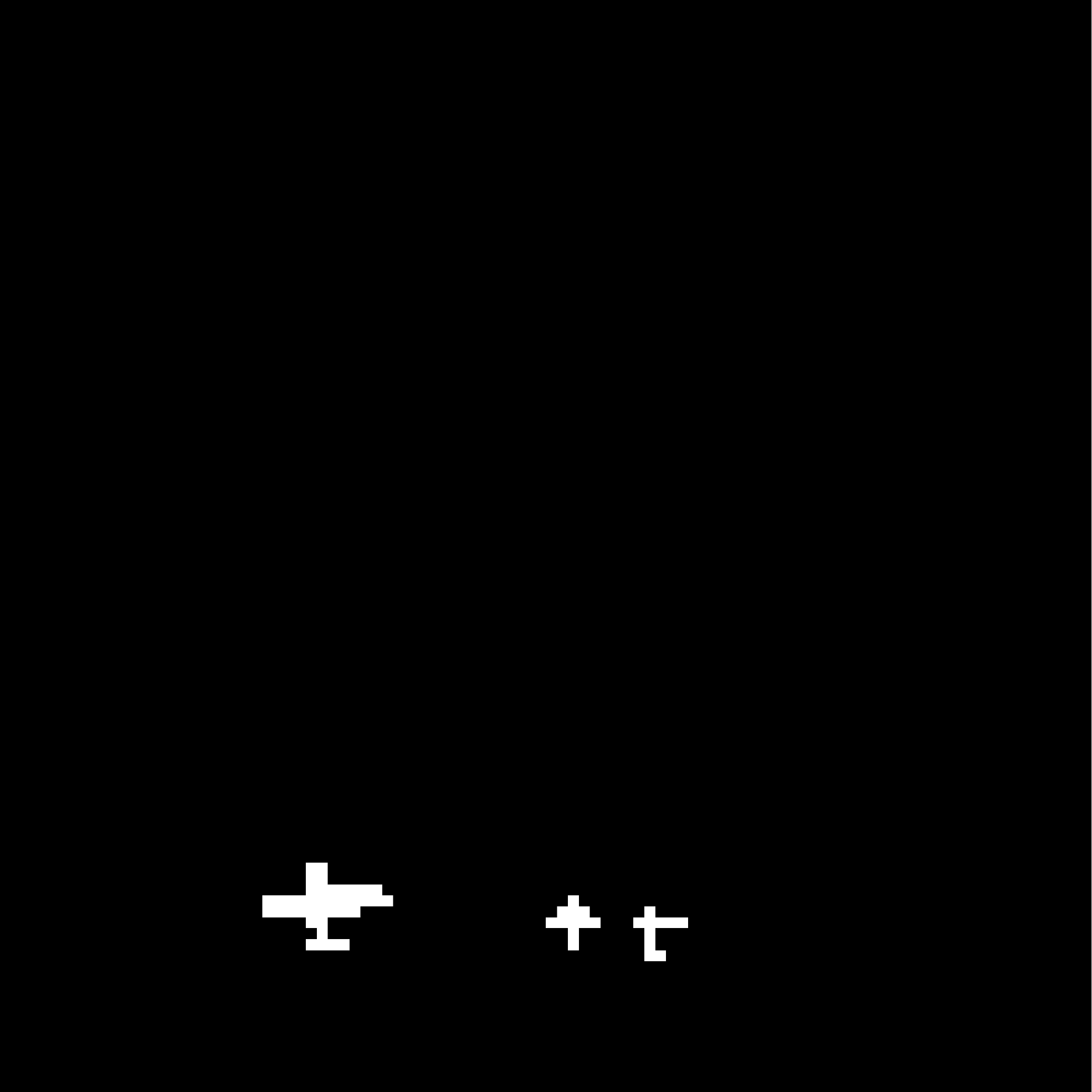}}
	\subfigure[\scriptsize{}]{
		\label{Fig:subfig:PaviaCentra-gt}
		\includegraphics[width=1.5 in]{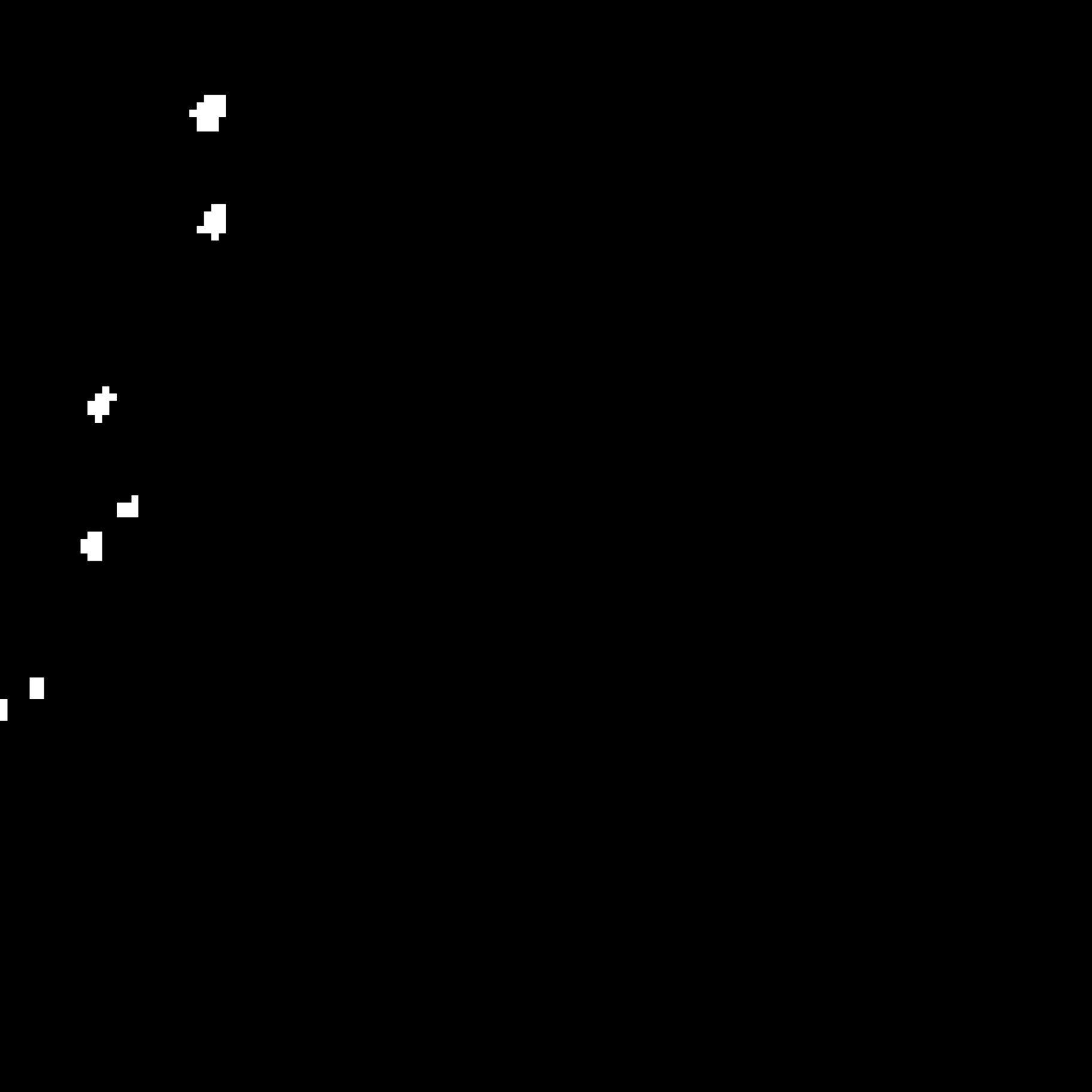}}
	\subfigure[\scriptsize{}]{
		\label{Fig:subfig:Gainesville-gt}
		\includegraphics[width=1.5 in]{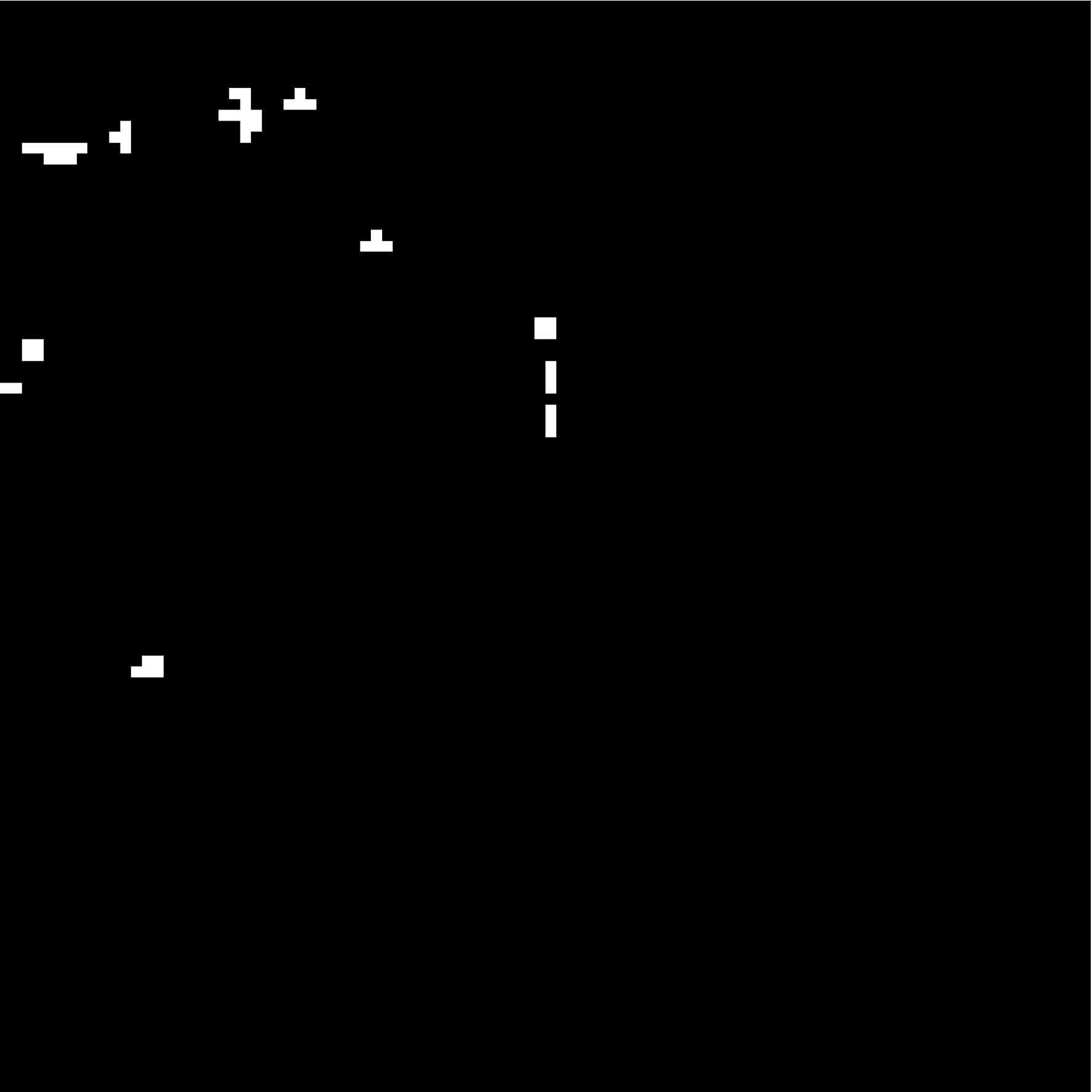}}
	\subfigure[\scriptsize{}]{
		\label{Fig:subfig:Cri-gt}
		\includegraphics[width=1.5 in]{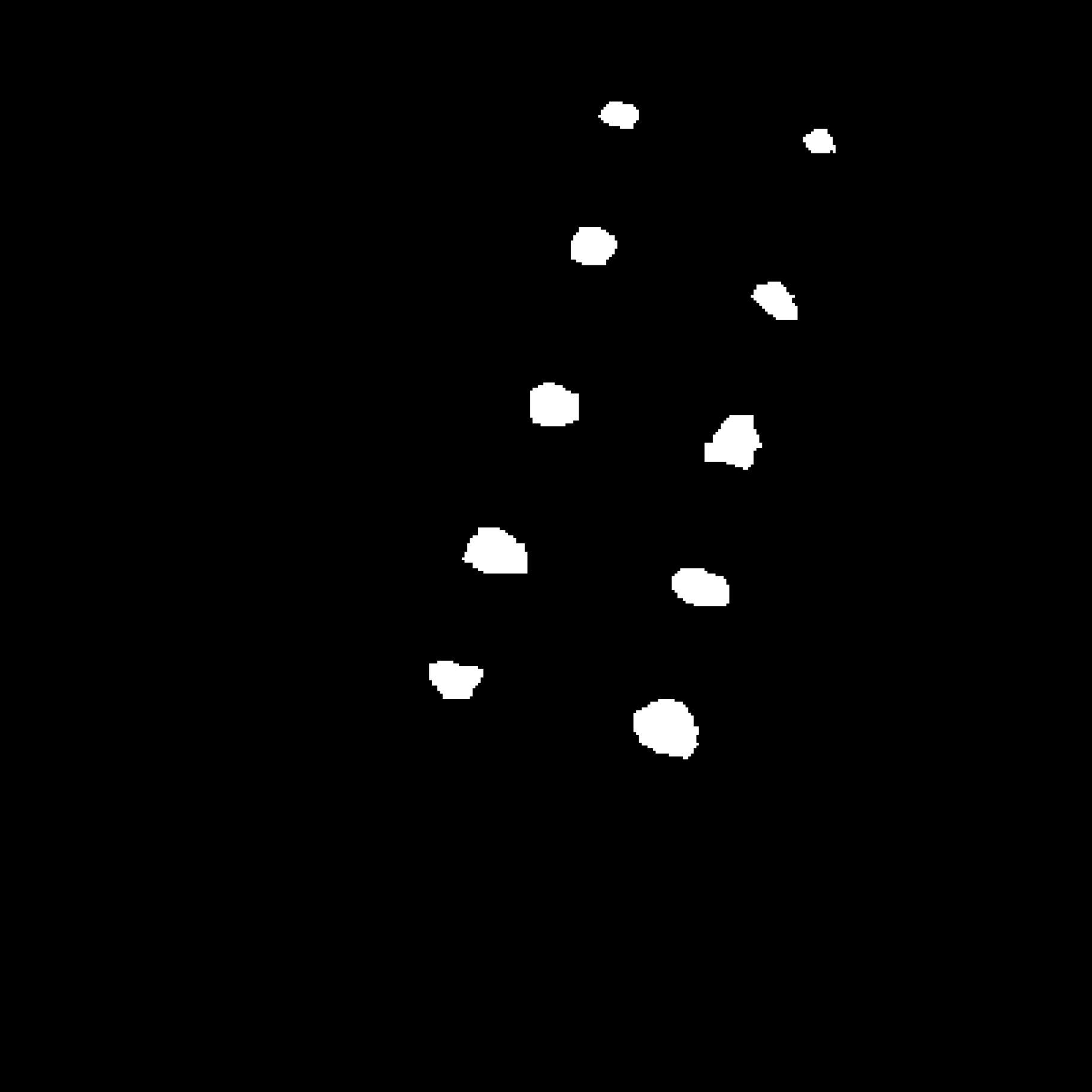}}
	\caption{The pseudo color presentation of (a) Gulfport, (b) Pavia Centra, (c) Gainesville, (d) Cri; the ground-truth map of (e) Gulfport, (f) Pavia Centra, (g) Gainesville, (h) Cri.}
	\label{datasets}
\end{figure*}

\section{Experiments and Discussion}

\subsection{Datasets Description}
In this section, to evaluate the performance of the proposed detector, experiments are conducted on four hyperspectral images. The first image \cite{kang2017hyperspectral} was obtained by airborne visible/infrared imaging spectrometer (AVIRIS) sensor over Gulfport area on July 7, 2010. This Gulfport image with spatial resolution of 3.4 m,  consists of 100 $\times$ 100 pixels and 191 spectral bands after removing the invalid bands, where the anomalies refer to the airplanes. The scenario and its ground-truth map are shown in Fig. \ref{Fig:subfig:Gulfport} and Fig.\ref{Fig:subfig:Gulfport-gt}. 

The second image \cite{kang2017hyperspectral,bitar2019sparse} was acquired by the Reflective Optics System Imaging Spectrometer (ROSIS) airborne sensor during a flight campaign over Pavia, northern Italy. Its size and geometric resolution are 150 $\times$ 150 $\times$ 115 and 1.3 m, respectively. 102 bands are retained as the noisy bands are discarded. The Pavia Centra image mainly contains a background of water and a bridge and some anomalies of vehicles, of which the scenario and the ground-truth map are given in Fig. \ref{Fig:subfig:PaviaCentra} and Fig. \ref{Fig:subfig:PaviaCentra-gt}. 

The third dataset \cite{kang2017hyperspectral} of spatial resolution 3.5 m is an urban scene of Gainesville located on the north central of Florida, USA, obtained by AVIRIS on September 4, 2010. It consists of 191 image bands of size 100 $\times$ 100. Fig. \ref{Fig:subfig:Gainesville} and Fig. \ref{Fig:subfig:Gainesville-gt} depict the Gainesville image and its corresponding ground truth, respectively.

The fourth dataset \cite{huyan2018hyperspectral,zhang2015low} was acquired by the Nuance Cri Hyperspectral sensor at the spectral resolution of 10 m and in the wavelength range of [650, 1100] nm. The Cri image of size 400 $\times$ 400 $\times$ 46 captures a grassland corrupted by 10 rock anomalies of pixel 2261. The pseudo color image is shown in Fig. \ref{Fig:subfig:Cri}, while the contained anomalies can be clearly observed in Fig. \ref{Fig:subfig:Cri-gt}.

\begin{figure*}[tp]
	\centering
	\subfigure[\scriptsize{}]{
		\label{Box_LosAngeles}
		\includegraphics[width=3 in]{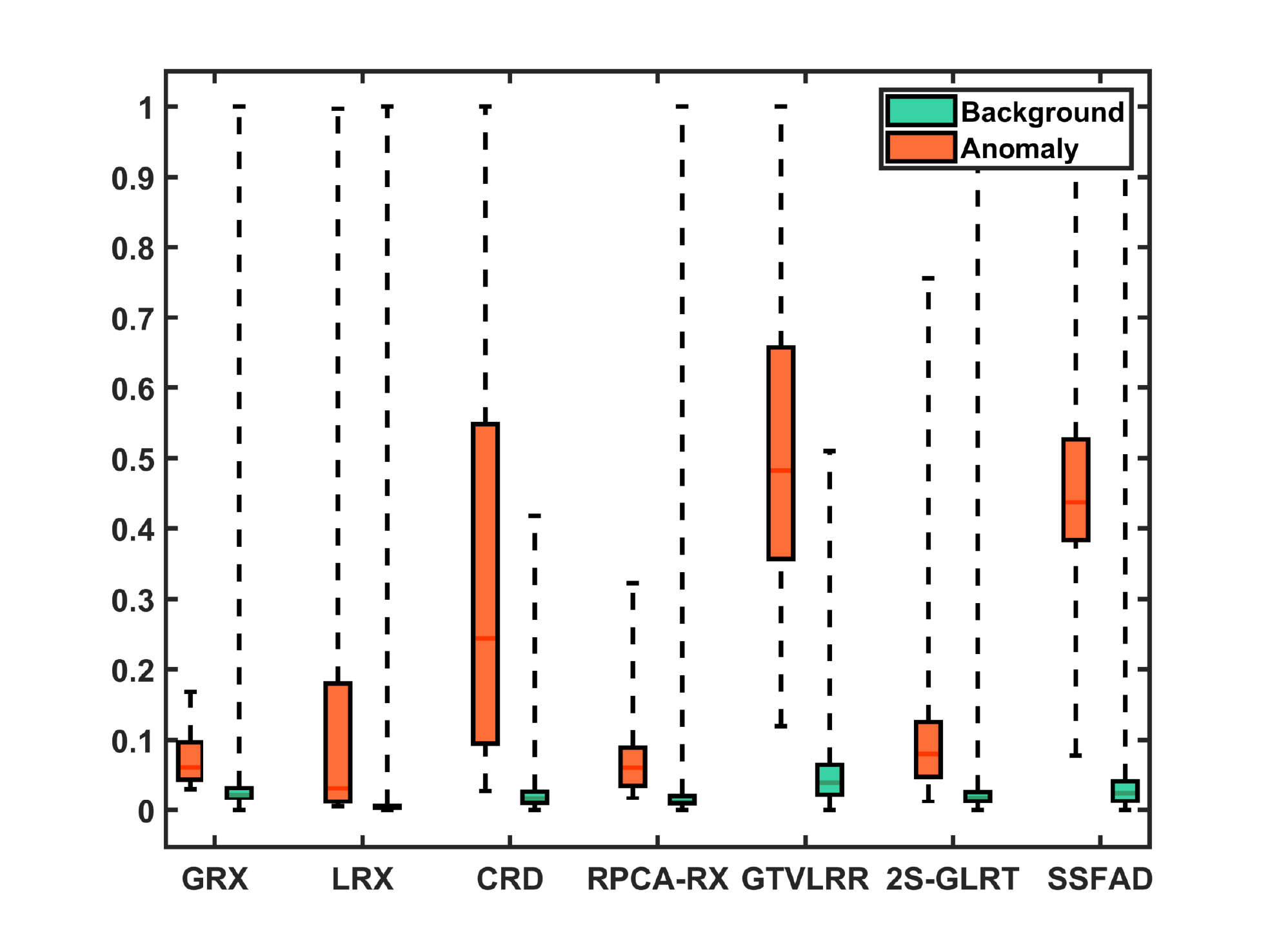}}
	\subfigure[\scriptsize{}]{
		\label{Box_CatIsland}
		\includegraphics[width=3 in]{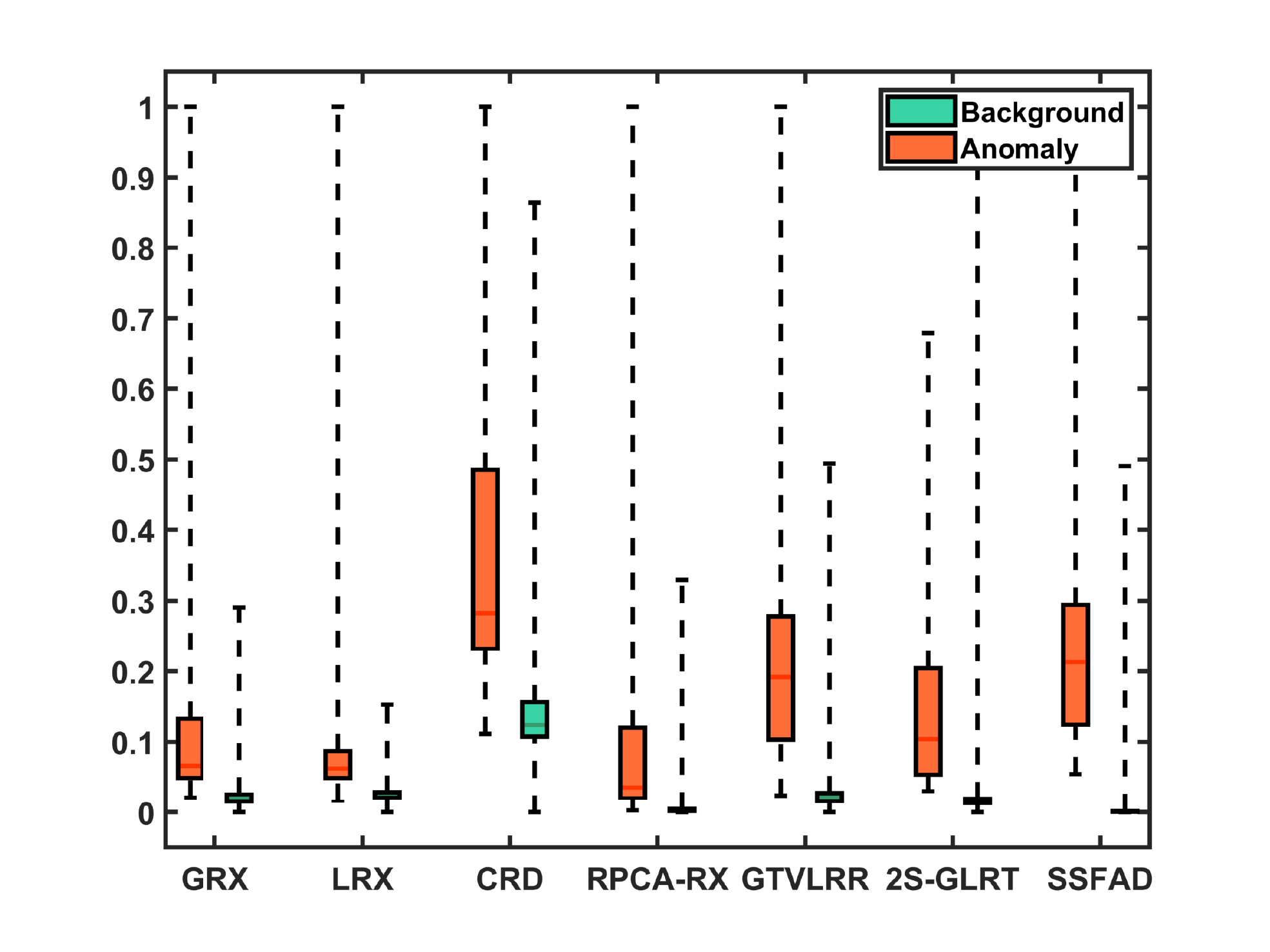}}
	\subfigure[\scriptsize{}]{
		\label{Box_PaviaCentra}
		\includegraphics[width=3 in]{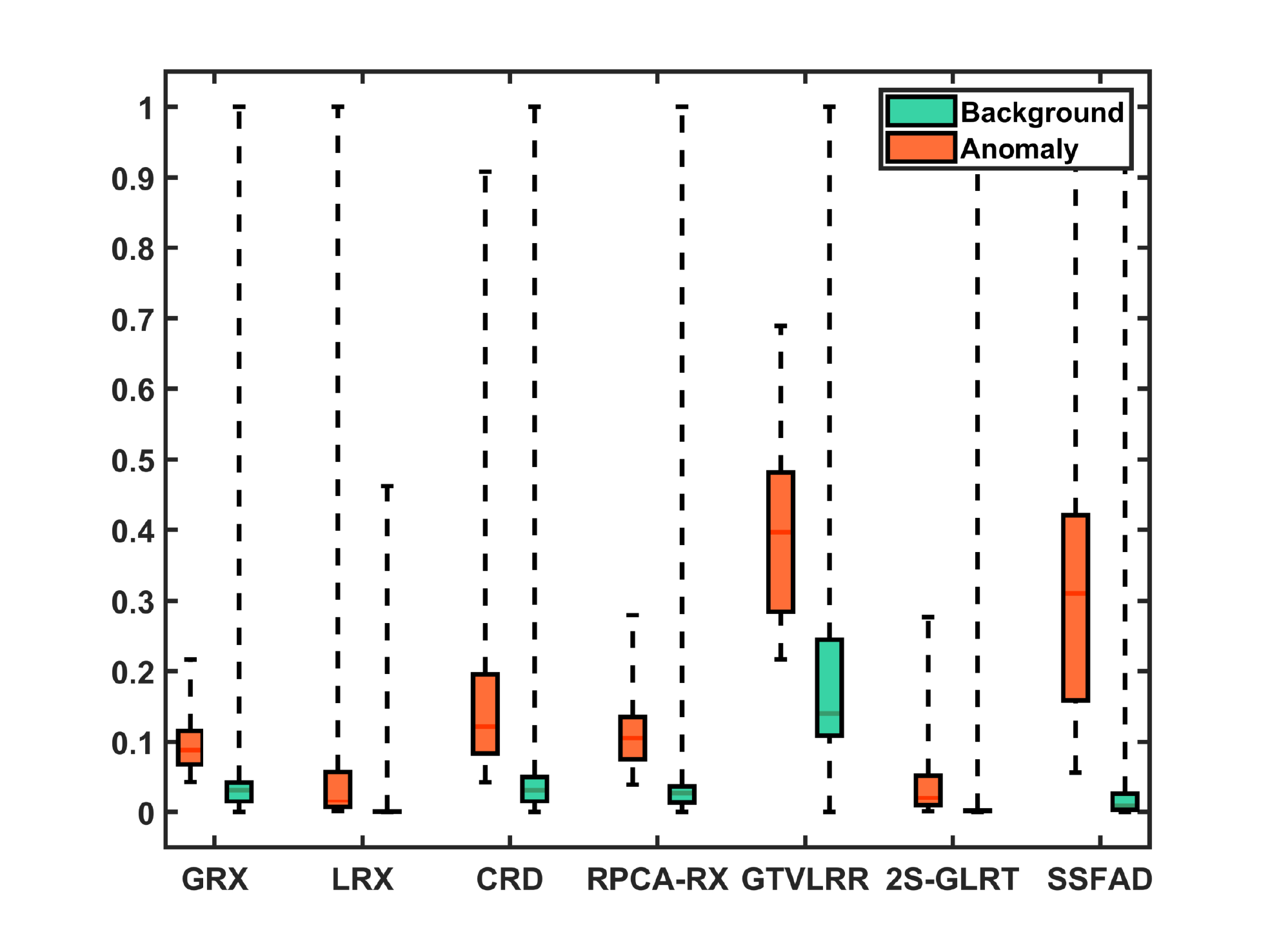}}
	\subfigure[\scriptsize{}]{	
		\label{Box_TexasCoast}
		\includegraphics[width=3 in]{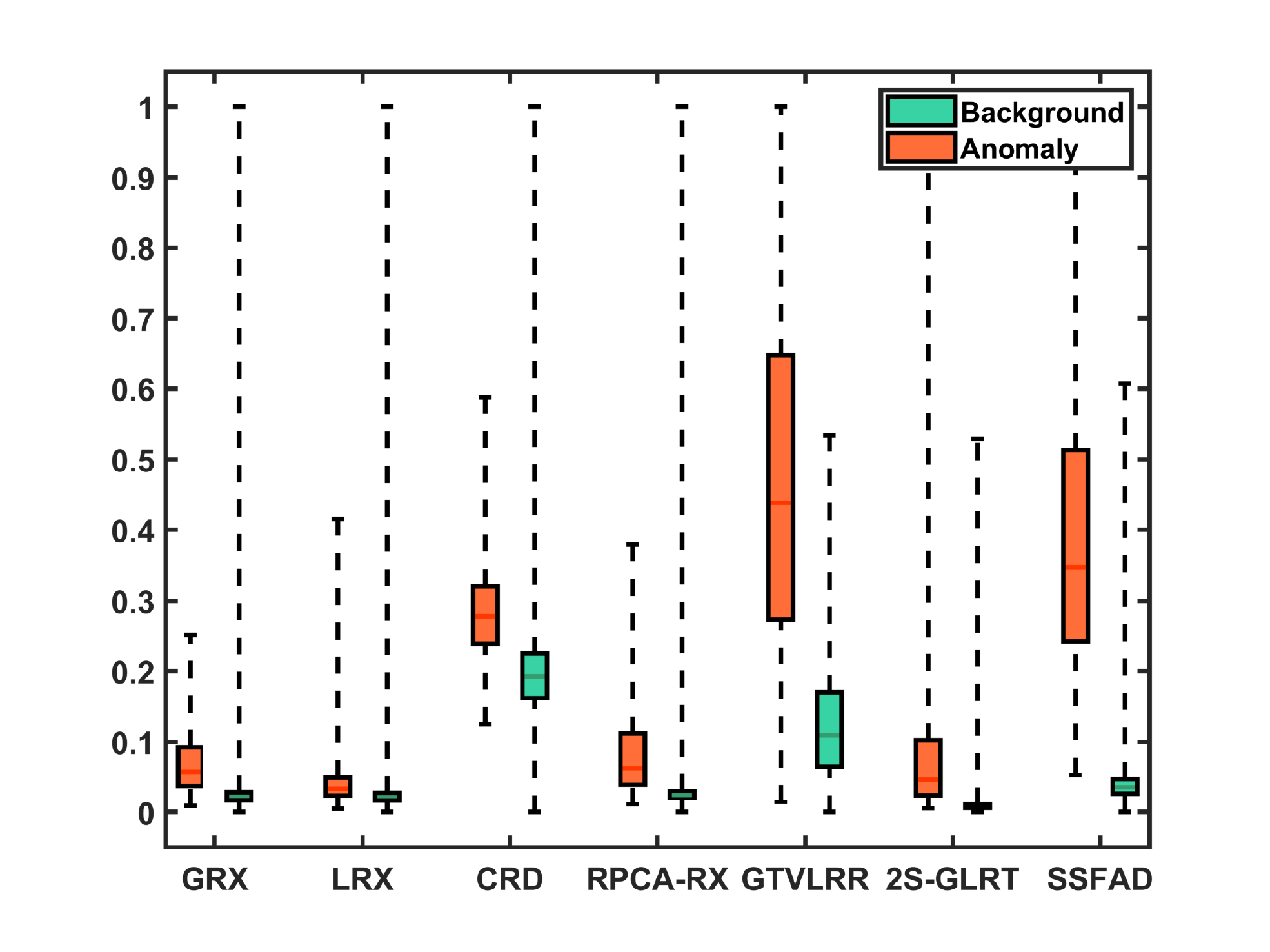}}
	\caption{Statistical separability analysis for different datasets: (a) Gulfport; (b) Pavia Centra; (c) Gainesville; (d) Cri.}
	\label{Boxplots}
\end{figure*}

\begin{figure*}[tp]
	\centering
	\subfigure[\scriptsize{}]{
		\label{LosAngeles_Results}
		\includegraphics[width=3 in]{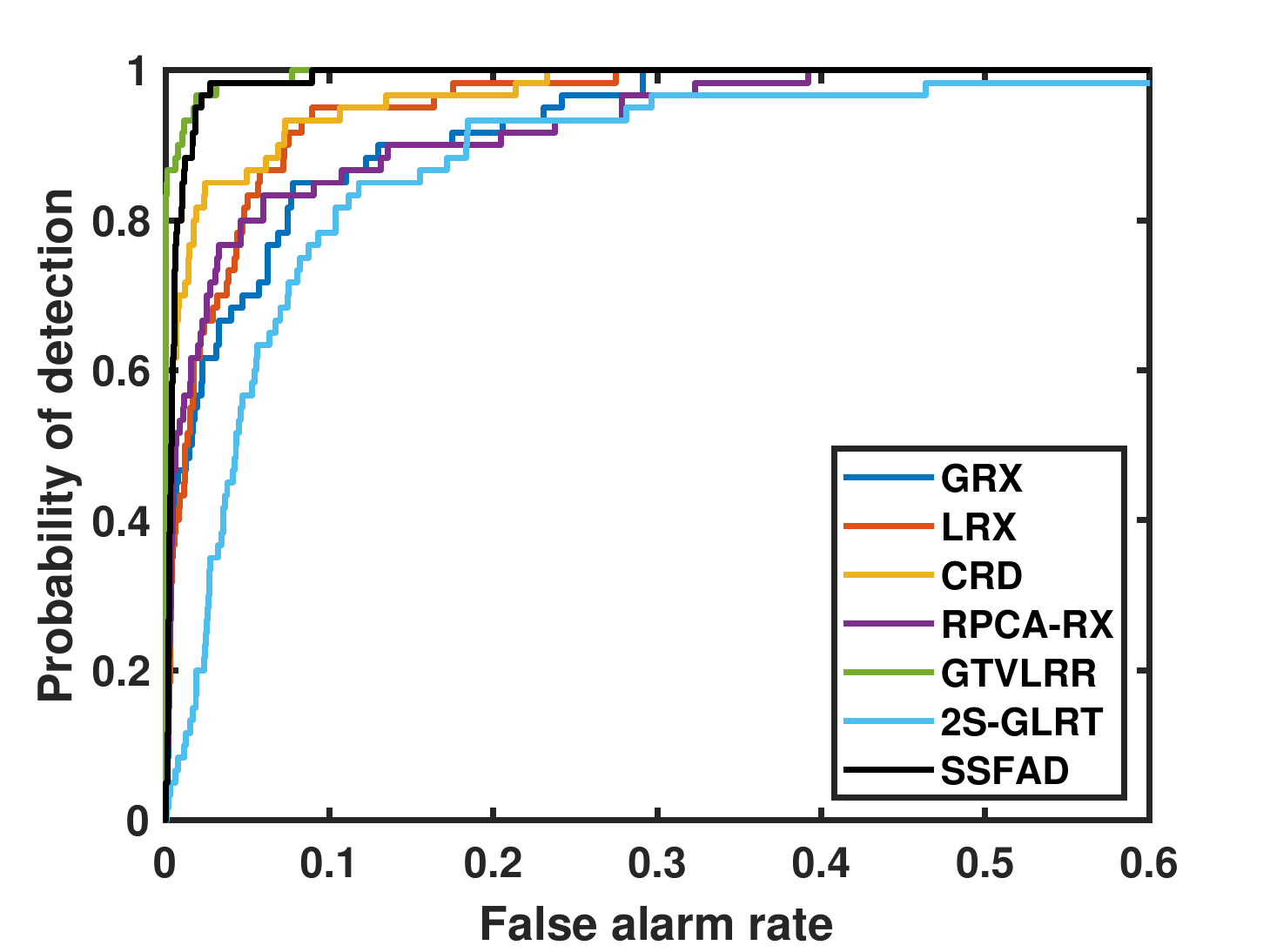}}
	\subfigure[\scriptsize{}]{
		\label{CatIsland_Results}
		\includegraphics[width=3 in]{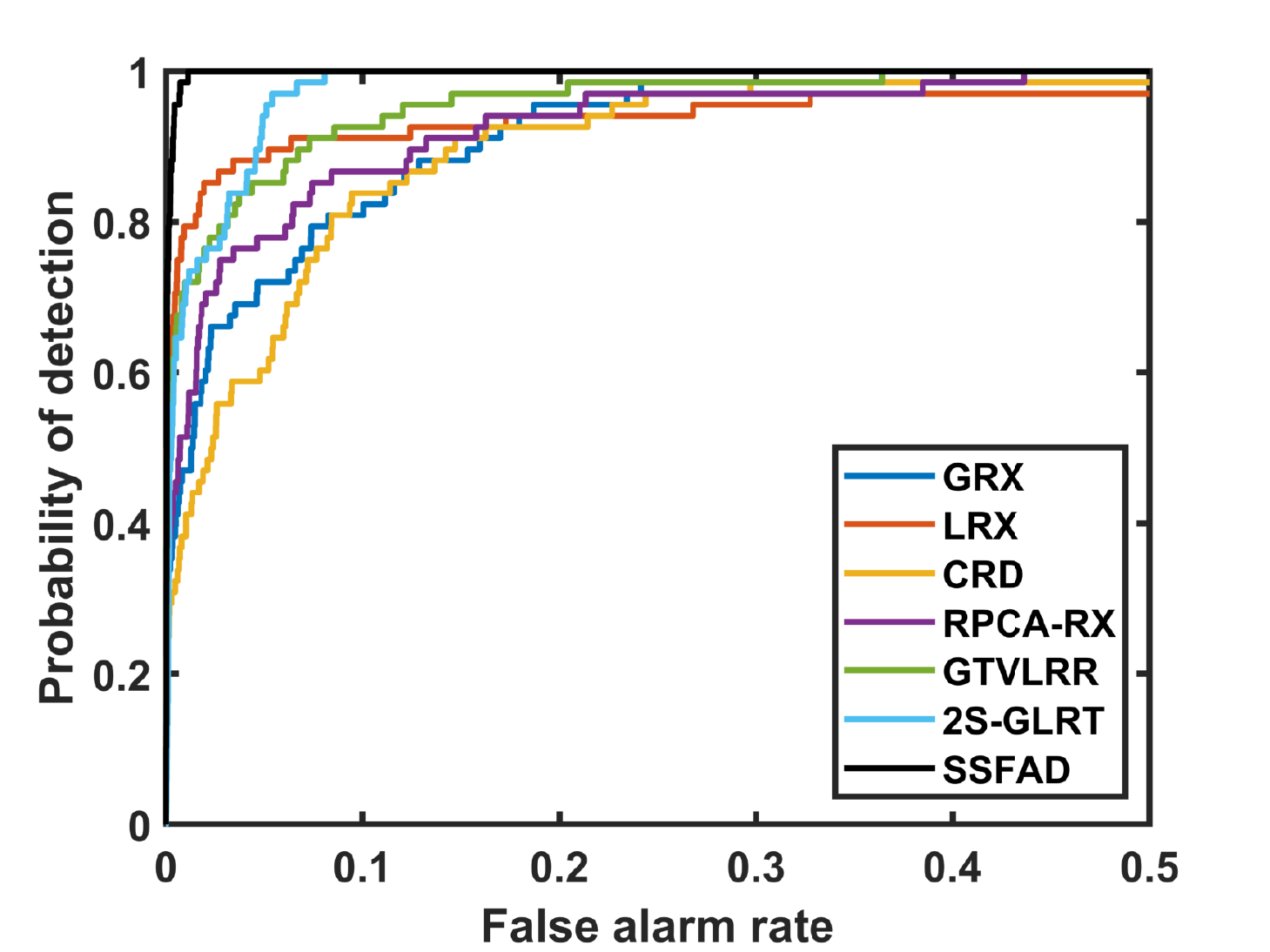}}
	\subfigure[\scriptsize{}]{
		\label{PaviaCentra_Results}
		\includegraphics[width=3 in]{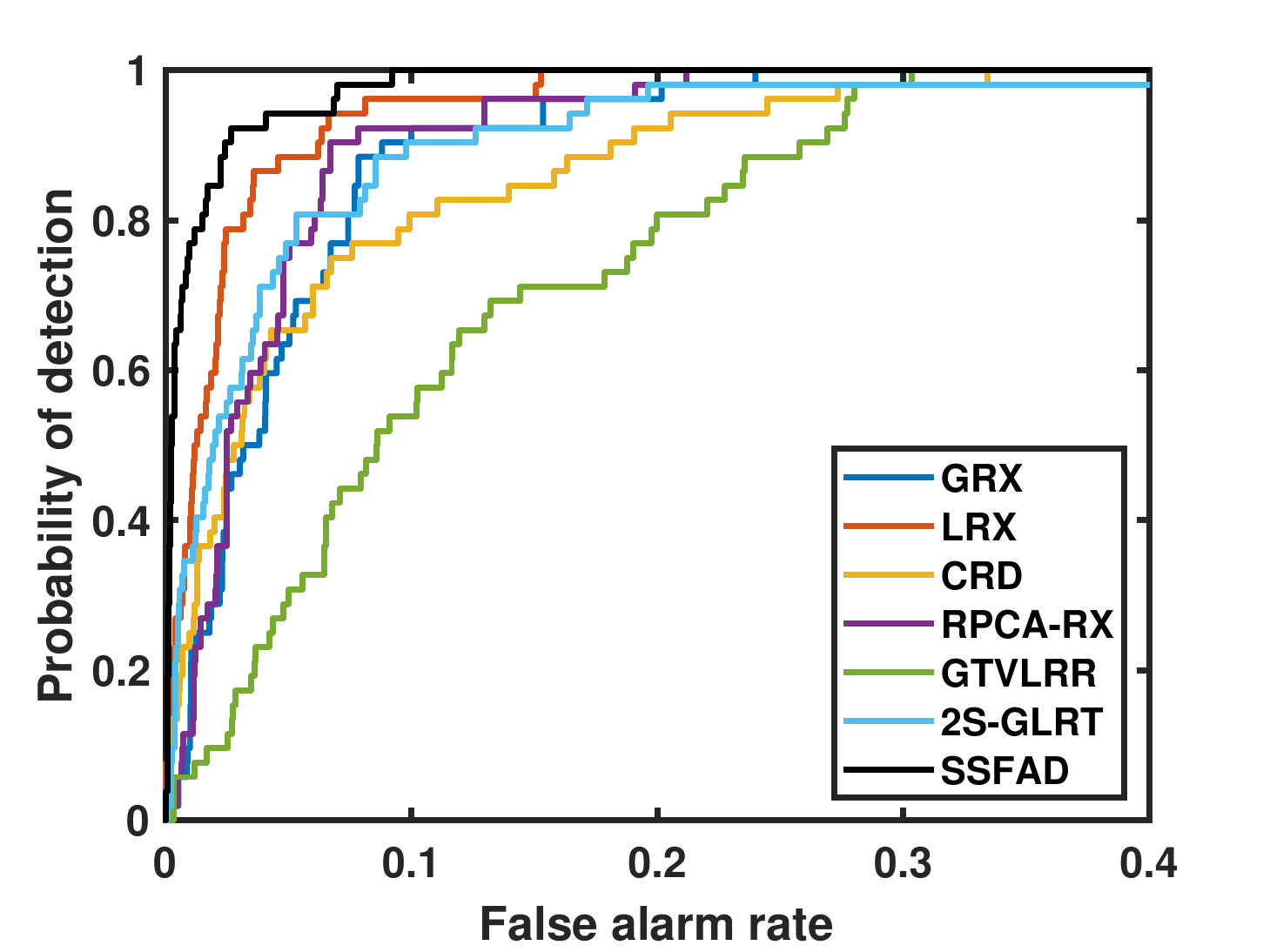}}
	\subfigure[\scriptsize{}]{
		\label{TexasCoast_Results}
		\includegraphics[width=3 in]{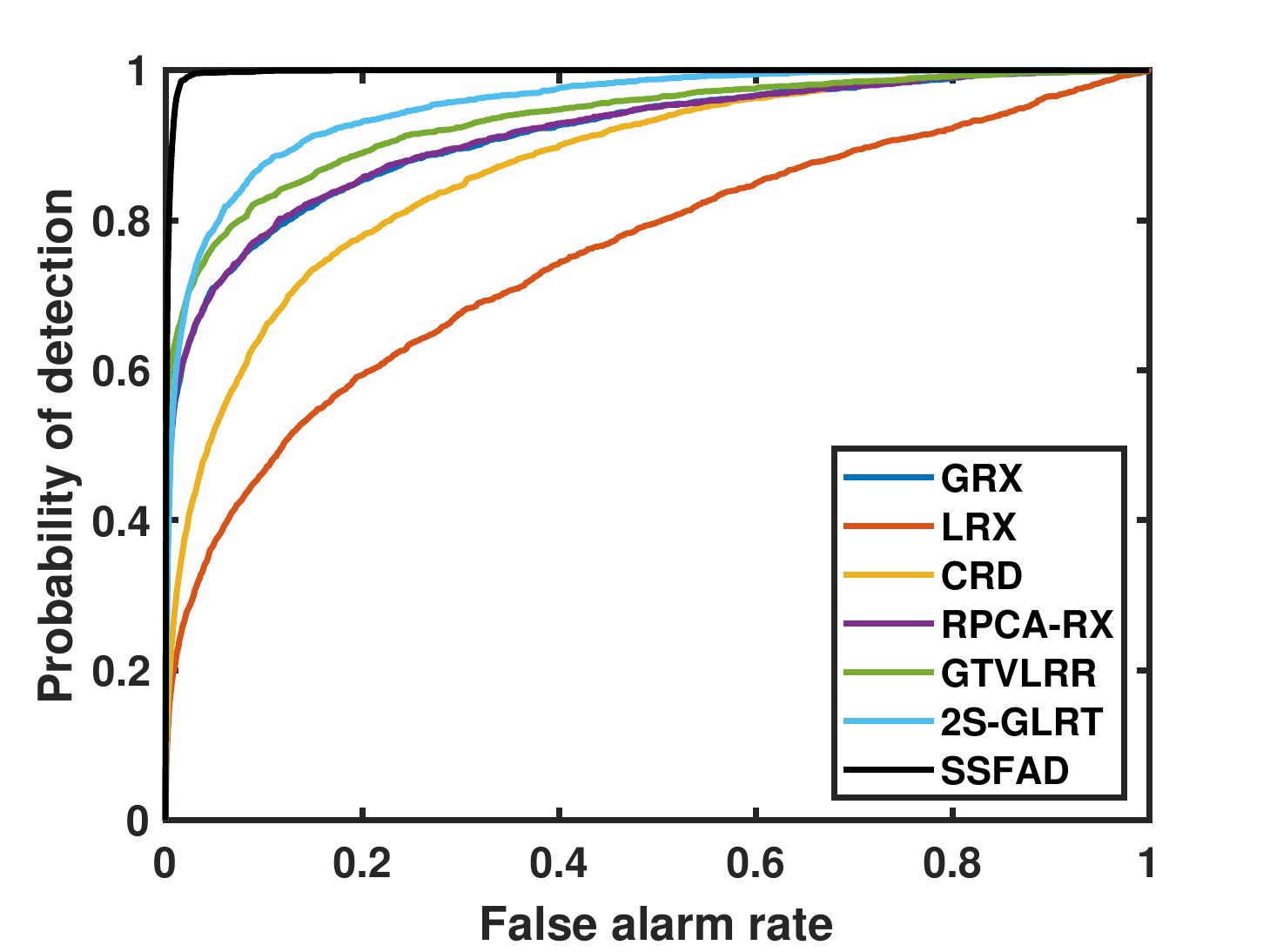}}
	\caption{ROC curves for different datasets: (a) Gulfport; (b) Pavia Centra; (c) Gainesville; (d) Cri.}
	\label{ROC_Values}
	
\end{figure*}

\begin{table}  
		\footnotesize
		\centering
		\caption{The optimal parameters of various detectors in different datasets.}
		\begin{tabular}{ccccc}
			\hline
			Methods & Gulfport & Pavia Centra & Gainesville  & Cri  \\
			\hline
			LRX	     &(17,15)  &(25,5)   &(9,7)   &(19,15) \\
			CRD	     &(23,15,$10^{-6}$)   &(7,5,$10^{-6}$)  &(9,7,$10^{-6}$)  &(19,9,$10^{-6}$) \\
			RPCA-RX	 &0.01     &0.01      &0.01    &0.001 \\
			GTVLRR  &(0.5,0.2,0.1) &(0.7,0.4,0.05) &(0.005,1,0.5)  &(0.005,1,0.5) \\
			2S-GLRT  &(13,11)  &(21,5)    &(9,5)  &(25,15) \\
			SSFAD    &(5,3)    &(5,3)     &(5,3)   &(17,15) \\
			\hline
		\end{tabular}
		\label{Tab:Optimal_Parameters}
	\end{table} 

\begin{table}
	\footnotesize
	\centering
	\caption{AUC values (\%) of different fusion strategies in SSFAD. The maximum value is highlighted by red}
	\begin{tabular}{cccccc}
		\hline
		Methods & Gulfport & Pavia Centra & Gainesville  & Cri   \\
		\hline
		average pooling & \bf{\textcolor{red}{99.309}}   & 99.843      & 98.903       & 99.614\\
		adaptive score  & 99.307        & \bf{\textcolor{red}{99.882}} & \bf{\textcolor{red}{98.970}}  & \bf{\textcolor{red}{99.751}}\\
		
		\hline
	\end{tabular}
	\label{Tab:fusion_strategy}
\end{table}

\subsection{Parameters setting}
To effectively verify the performance of the proposed SSFAD method, six widely-used anomaly detectors, e.g. GRX, LRX, CRD, RPCA-RX, GTVLRR and 2S-GLRT, are employed for evaluation. Meanwhile, for qualitative and quantitative comparisons, statistical separability analysis \cite{houcollaborative}, receiver operating characteristic (ROC) \cite{hanley1982meaning} and area under the curve (AUC) metric are utilized as the main criteria.

The choices of parameters are important for these methods with parameters, by which its detection performance is greatly affected. For a fair comparison, the parameters of all considered methods are carefully tuned following their expriences, to generate their best detection results. As suggested in \cite{liu2021multipixel,houcollaborative}, for the LRX and SSFAD, by varying the outer window size $\omega_{\text{out}}$ from 5 to 25 and the inner window size $\omega_{\text{in}}$ from 3 to 15, the optimal detection performance under different window sizes $(\omega_{\text{out}}, \omega_{\text{in}})$ is collected. For the CRD, by fixing $\lambda$ as $10^{-6}$ and varying $\omega_{\text{out}}$ from 5 to 25 and $\omega_{\text{in}}$ from 3 to 15, the optimal detection performance under different window sizes is selected. For the RPCA-RX, the optimal parameter $\lambda$ is selected from $10^{-4}$ to $10^{4}$ for comparative experiments. As to the GTVLRR, the K-means method is used to construct dictionary referring to \cite{cheng2019graph,liu2021multipixel}, and empirical parameters ($\lambda$, $\beta$, $\gamma$) are firstly set to (0.5, 0.2, 0.05). Then, by fixing $\beta$ and $\gamma$, and varying  $\lambda$ in [0.005, 0.05, 0.1, 0.3, 0.5, 0.7, 1], the optimal $\lambda$ under different values is selected. Similarly, by fixing the selected $\lambda$ and default $\gamma$, and varying  $\beta$ in [0.005, 0.05, 0.1, 0.2, 0.4, 0.7, 1], the optimal $\beta$ is selected. In the same way, the optimal $\gamma$ is selected in [0.005, 0.01, 0.02, 0.05, 0.1, 0.2, 0.5]. The optimal parameters of various detectors are listed in Table \ref{Tab:Optimal_Parameters}.

\begin{table}
	\footnotesize
	\centering
	\caption{AUC values (\%) of various anomaly detectors using different datasets. The top two values are highlighted by red and blue}
	\begin{tabular}{cccccc}
		\hline
		Methods & Gulfport & Pavia Centra & Gainesville  & Cri   \\
		\hline
		GRX      & 95.260      & 95.380       & 95.128       & 91.342 \\
		LRX	     & 96.942      & 95.973       & \bf{\textcolor{blue}{97.690}}       & 75.251 \\
		CRD	     & 97.927      & 94.072       & 93.930       & 87.267 \\
		RPCA-RX	 & 95.348      & 95.906       & 95.861       & 91.478 \\
		GTVLRR	 & \bf{\textcolor{red}{99.691}}      & 97.636       & 88.604       & 93.429 \\
		2S-GLRT	 & 91.831      & \bf{\textcolor{blue}{98.676}}       & 94.856       & \bf{\textcolor{blue}{95.403}} \\
		SSFAD    & \bf{\textcolor{blue}{99.307}}      & \bf{\textcolor{red}{99.882}}       & \bf{\textcolor{red}{98.970}}       & \bf{\textcolor{red}{99.751}}\\
		
		\hline
	\end{tabular}
	\label{Tab:AUC_Values}
\end{table}

\begin{table}
	\footnotesize
	\centering
	\caption{Execution time of various anomaly detectors using different datasets (unit: seconds).}
	\begin{tabular}{cccccc}
		\hline
		Methods & Gulfport & Pavia Centra & Gainesville  & Cri   \\
		\hline
		GRX     & 0.08     & 0.08      & 0.12     & 0.32      \\
		LRX	    & 39.57    & 50.98     & 34.14    & 91.59     \\
		CRD	    & 69.97    & 4.79      & 3.87     & 866.59    \\
		RPCA-RX	& 2.61     & 2.19      & 2.56     & 7.76      \\
		GTVLRR	& 61.06    & 88.89     & 49.40    & 981.47    \\
		2S-GLRT	& 74.36    & 48.11     & 36.60    & 4272.53   \\
		SSFAD   & 50.44    & 44.67     & 47.67    & 3197.97   \\
		\hline
	\end{tabular}
	\label{Tab:execution_time}
\end{table}

\subsection{Detection Performance}
Statistical separability analysis \cite{liu2021multipixel,zhao2021hyperspectral}, also known as boxplot, has been widely used as a performance assessment in hyperspectral anomaly detection, target detection, and change detection. Essentially, the boxplot is used to reflect the distribution characteristics of various detection results, where the separation degree between different categories is judged by the interval between different boxes. Herein, the separability and the suppression level between anomaly and background is evaluated by statistical separability analysis, where the red and green boxes represent the anomalous and background pixels, respectively. The interval between the red and green box represents the separability between the anomalous and background pixels, while the height of the green box means the suppression degree of these methods to the background. Generally, a shorter green box corresponds to a stronger background suppression, refelcting a better separation for background and anomaly. 

The detection performance of the considered algorithms on four experimental datasets is displayed. As can be seen from Fig. \ref{Boxplots}(a) that the intervals between the red and green boxes of SSFAD and GTVLRR are larger than those of the GRX, LRX, CRD, RPCA-RX and 2S-GLRT, which indicates that both methods can separate the anomalies well from the background. In addition, SSFAD is sightly better than GTVLRR towards interval, and perform the shortest green box. Thus, this proposed method can suppress the background better in the Gulfport dataset. Fig. \ref{Boxplots}(b) shows that except CRD, the compared detectors can effectively suppress the background of the Pavia Centra dataset. Obviously, SSFAD achieves the best separability than the others. From Fig. \ref{Boxplots}(c) and Fig. \ref{Boxplots}(d), it can be seen that SSFAD produces the widest interval when compared with other six detectors, i.e., GRX, LRX, CRD, RPCA-RX, GTVLRR and 2S-GLRT. Comprehensively,  the proposed SSFAD can separate anomalies from background more effectively. 

For an accuracy evaluation, the ROC curves of each detector on the experimental images are plotted in Fig. \ref{ROC_Values}. The closer the curve is to the upper left corner, the better the detection performance. As Fig. \ref{ROC_Values}(a) depicted, the GRX shows the worst performance. The green and black curves are closer to the upper left corner than others, representing that GTVLRR and the proposed SSFAD both achiev good detection performance, and GTVLRR is slightly better than SSFAD. The experimental results shown in Figs. \ref{ROC_Values}(b)-(d) reveal the obvious advantage of the proposed SSFAD when detecting the Pavia Centra, Gainesville, and Cri dataset, because the black curve is closest to the upper left corner. 

The AUC values of different fusion strategies are compared in Table \ref{Tab:fusion_strategy}. Obviously, the proposed adaptive score strategy achieves better results than the average pooling strategy, especially in Cri dataset. This also further confirms that the adaptive score stategy can better highlight anomaly than the average pooling strategy. In addition, the AUC values of various methods on differet datasets are listed in Table \ref{Tab:AUC_Values}. By comparing these AUC values, the almost best performance of proposed SSFAD can be observed. Especially on the Cri dataset, the improvement of SSFAD approximates to 4\% when compared with the second best detector. Table \ref{Tab:execution_time} provides the execution time of the compared methods, where the mediocre time performance of the proposed SSFAD is exposed. Such unideal time consuming probably is caused by the computational cost of matrix inversion operation and saliency weight calculation increasing with the size of the additional local window.  Whatever, the ordinary time performance can hardly cover the superiorities of the proposed SSFAD towards hyperspectaral anomaly detection.

\section{Conclusions}
In this paper, a novel dual-pipeline framework that extracted spatial and spectral features separately and then fused them adaptively was proposed for hyperspectral anomaly detection. In spectral domain, original spectral signal was first mapped to a local median-mean linear background, and then saliency weight and feature enhancement strategies were implemented to obtain an initial detection map. In spatial domain, a new detector was designed to fully excavate the similarity spatial information of local patches around center window. Finally, anomalies were detected by jointly considering the spectral and spatial detection maps. Experimental results on four real hyperspectral datasets confirmed that the proposed SSFAD method had superior background suppression ability and excellent detection performance than traditional anomaly detection methods.

\bibliographystyle{unsrtnat}
\bibliography{bibHou}  

\begin{thebibliography}{40}
\providecommand{\natexlab}[1]{#1}
\providecommand{\url}[1]{\texttt{#1}}
\expandafter\ifx\csname urlstyle\endcsname\relax
  \providecommand{\doi}[1]{doi: #1}\else
  \providecommand{\doi}{doi: \begingroup \urlstyle{rm}\Url}\fi

\bibitem[Hou et~al.(2018)Hou, Chen, Tan, and Du]{hou2018novel}
Zengfu Hou, Yu~Chen, Kun Tan, and Peijun Du.
\newblock Novel hyperspectral anomaly detection methods based on unsupervised
  nearest regularized subspace.
\newblock \emph{International Archives of the Photogrammetry, Remote Sensing \&
  Spatial Information Sciences}, 42\penalty0 (3), 2018.

\bibitem[Hou and Li(2022)]{hou2022joint}
Zengfu Hou and Wei Li.
\newblock A joint morphological profiles and patch tensor change detection for
  hyperspectral imagery.
\newblock \emph{arXiv preprint arXiv:2201.08027}, 2022.

\bibitem[Zhao et~al.(2021)Zhao, Hou, Wu, Li, Ma, and
  Tao]{zhao2021hyperspectral}
Xiaobin Zhao, Zengfu Hou, Xin Wu, Wei Li, Pengge Ma, and Ran Tao.
\newblock Hyperspectral target detection based on transform domain adaptive
  constrained energy minimization.
\newblock \emph{International Journal of Applied Earth Observation and
  Geoinformation}, 103:\penalty0 102461, 2021.

\bibitem[Hu et~al.(2017)Hu, Mou, Schmitt, and Zhu]{hu2017fusionet}
Jingliang Hu, Lichao Mou, Andreas Schmitt, and Xiao~Xiang Zhu.
\newblock Fusionet: A two-stream convolutional neural network for urban scene
  classification using polsar and hyperspectral data.
\newblock In \emph{2017 Joint Urban Remote Sensing Event (JURSE)}, pages 1--4.
  IEEE, 2017.

\bibitem[Villa et~al.(2010)Villa, Chanussot, Benediktsson, and
  Jutten]{villa2010spectral}
Alberto Villa, Jocelyn Chanussot, Jon~Atli Benediktsson, and Christian Jutten.
\newblock Spectral unmixing for the classification of hyperspectral images at a
  finer spatial resolution.
\newblock \emph{IEEE Journal of Selected Topics in Signal Processing},
  5\penalty0 (3):\penalty0 521--533, 2010.

\bibitem[Hou et~al.(2021{\natexlab{a}})Hou, Li, Li, Tao, and
  Du]{hou2021hyperspectral}
Zengfu Hou, Wei Li, Lu~Li, Ran Tao, and Qian Du.
\newblock Hyperspectral change detection based on multiple morphological
  profiles.
\newblock \emph{IEEE Transactions on Geoscience and Remote Sensing},
  2021{\natexlab{a}}.

\bibitem[Hou et~al.(2021{\natexlab{b}})Hou, Li, Tao, and Du]{hou2021three}
Zengfu Hou, Wei Li, Ran Tao, and Qian Du.
\newblock Three-order tucker decomposition and reconstruction detector for
  unsupervised hyperspectral change detection.
\newblock \emph{IEEE Journal of Selected Topics in Applied Earth Observations
  and Remote Sensing}, 14:\penalty0 6194--6205, 2021{\natexlab{b}}.

\bibitem[Liu et~al.(2021)Liu, Hou, Li, Tao, Orlando, and Li]{liu2021multipixel}
Jun Liu, Zengfu Hou, Wei Li, Ran Tao, Danilo Orlando, and Hongbin Li.
\newblock Multipixel anomaly detection with unknown patterns for hyperspectral
  imagery.
\newblock \emph{IEEE Transactions on Neural Networks and Learning Systems},
  pages 1--11, 2021.
\newblock \doi{10.1109/TNNLS.2021.3071026}.

\bibitem[Hou et~al.(2020)Hou, Li, Gao, Zhang, Ma, and Sun]{hou2020background}
Zengfu Hou, Wei Li, Lianru Gao, Bing Zhang, Pengge Ma, and Junling Sun.
\newblock A background refinement collaborative representation method with
  saliency weight for hyperspectral anomaly detection.
\newblock In \emph{IGARSS 2020-2020 IEEE International Geoscience and Remote
  Sensing Symposium}, pages 2412--2415. IEEE, 2020.

\bibitem[Hou et~al.(2021{\natexlab{c}})Hou, Wei, Tao, and
  Shi]{houcollaborative}
Zengfu Hou, Li~Wei, Ran Tao, and Weihua Shi.
\newblock Collaborative representation with background purification and
  saliency weight for hyperspectral anomaly detection.
\newblock \emph{SCIENCE CHINA Information Sciences}, 65:\penalty0 112305,
  2021{\natexlab{c}}.

\bibitem[Reed and Yu(1990)]{reed1990adaptive}
Irving~S Reed and Xiaoli Yu.
\newblock Adaptive multiple-band cfar detection of an optical pattern with
  unknown spectral distribution.
\newblock \emph{IEEE Transactions on Acoustics, Speech, and Signal Processing},
  38\penalty0 (10):\penalty0 1760--1770, 1990.

\bibitem[Molero et~al.(2013)Molero, Garz{\'o}n, Garc{\'\i}a, and
  Plaza]{molero2013analysis}
Jos{\'e}~Manuel Molero, Ester~M Garz{\'o}n, Inmaculada Garc{\'\i}a, and Antonio
  Plaza.
\newblock Analysis and optimizations of global and local versions of the rx
  algorithm for anomaly detection in hyperspectral data.
\newblock \emph{IEEE Journal of Selected Topics in Applied Earth Observations
  and Remote Sensing}, 6\penalty0 (2):\penalty0 801--814, 2013.

\bibitem[Guo et~al.(2014)Guo, Zhang, Ran, Gao, Li, and Plaza]{guo2014weighted}
Qiandong Guo, Bing Zhang, Qiong Ran, Lianru Gao, Jun Li, and Antonio Plaza.
\newblock Weighted-rxd and linear filter-based rxd: Improving background
  statistics estimation for anomaly detection in hyperspectral imagery.
\newblock \emph{IEEE Journal of Selected Topics in Applied Earth Observations
  and Remote Sensing}, 7\penalty0 (6):\penalty0 2351--2366, 2014.

\bibitem[Kwon and Nasrabadi(2005)]{kwon2005kernel}
Heesung Kwon and Nasser~M Nasrabadi.
\newblock Kernel rx-algorithm: A nonlinear anomaly detector for hyperspectral
  imagery.
\newblock \emph{IEEE transactions on Geoscience and Remote Sensing},
  43\penalty0 (2):\penalty0 388--397, 2005.

\bibitem[Li and Du(2014)]{li2014collaborative}
Wei Li and Qian Du.
\newblock Collaborative representation for hyperspectral anomaly detection.
\newblock \emph{IEEE Transactions on Geoscience and Remote Sensing},
  53\penalty0 (3):\penalty0 1463--1474, 2014.

\bibitem[Vafadar and Ghassemian(2017)]{vafadar2017hyperspectral}
Maryam Vafadar and Hassan Ghassemian.
\newblock Hyperspectral anomaly detection using outlier removal from
  collaborative representation.
\newblock In \emph{2017 3rd International Conference on Pattern Recognition and
  Image Analysis (IPRIA)}, pages 13--19. IEEE, 2017.

\bibitem[Su et~al.(2018)Su, Wu, Du, and Du]{su2018hyperspectral}
Hongjun Su, Zhaoyue Wu, Qian Du, and Peijun Du.
\newblock Hyperspectral anomaly detection using collaborative representation
  with outlier removal.
\newblock \emph{IEEE Journal of Selected Topics in Applied Earth Observations
  and Remote Sensing}, 11\penalty0 (12):\penalty0 5029--5038, 2018.

\bibitem[Tan et~al.(2019{\natexlab{a}})Tan, Hou, Wu, Du, and
  Chen]{tan2019anomalysubspace}
Kun Tan, Zengfu Hou, Fuyu Wu, Qian Du, and Yu~Chen.
\newblock Anomaly detection for hyperspectral imagery based on the regularized
  subspace method and collaborative representation.
\newblock \emph{Remote sensing}, 11\penalty0 (11):\penalty0 1318,
  2019{\natexlab{a}}.

\bibitem[Cand{\`e}s et~al.(2011)Cand{\`e}s, Li, Ma, and
  Wright]{candes2011robust}
Emmanuel~J Cand{\`e}s, Xiaodong Li, Yi~Ma, and John Wright.
\newblock Robust principal component analysis?
\newblock \emph{Journal of the ACM (JACM)}, 58\penalty0 (3):\penalty0 1--37,
  2011.

\bibitem[Liu et~al.(2012)Liu, Lin, Yan, Sun, Yu, and Ma]{liu2012robust}
Guangcan Liu, Zhouchen Lin, Shuicheng Yan, Ju~Sun, Yong Yu, and Yi~Ma.
\newblock Robust recovery of subspace structures by low-rank representation.
\newblock \emph{IEEE transactions on pattern analysis and machine
  intelligence}, 35\penalty0 (1):\penalty0 171--184, 2012.

\bibitem[Xu et~al.(2015{\natexlab{a}})Xu, Wu, Wei, Liu, and Xu]{xu2015novel}
Yang Xu, Zebin Wu, Zhihui Wei, Hongyi Liu, and Xiong Xu.
\newblock A novel hyperspectral image anomaly detection method based on low
  rank representation.
\newblock In \emph{2015 IEEE International Geoscience and Remote Sensing
  Symposium (IGARSS)}, pages 4444--4447. IEEE, 2015{\natexlab{a}}.

\bibitem[Niu and Wang(2016)]{niu2016hyperspectral}
Yubin Niu and Bin Wang.
\newblock Hyperspectral anomaly detection based on low-rank representation and
  learned dictionary.
\newblock \emph{Remote Sensing}, 8\penalty0 (4):\penalty0 289, 2016.

\bibitem[Tan et~al.(2019{\natexlab{b}})Tan, Hou, Ma, Chen, and
  Du]{tan2019anomalylowrank}
Kun Tan, Zengfu Hou, Donglei Ma, Yu~Chen, and Qian Du.
\newblock Anomaly detection in hyperspectral imagery based on low-rank
  representation incorporating a spatial constraint.
\newblock \emph{Remote Sensing}, 11\penalty0 (13):\penalty0 1578,
  2019{\natexlab{b}}.

\bibitem[Chen et~al.(2011)Chen, Nasrabadi, and Tran]{chen2011simultaneous}
Yi~Chen, Nasser~M Nasrabadi, and Trac~D Tran.
\newblock Simultaneous joint sparsity model for target detection in
  hyperspectral imagery.
\newblock \emph{IEEE Geoscience and Remote Sensing Letters}, 8\penalty0
  (4):\penalty0 676--680, 2011.

\bibitem[Xu et~al.(2015{\natexlab{b}})Xu, Wu, Li, Plaza, and
  Wei]{xu2015anomaly}
Yang Xu, Zebin Wu, Jun Li, Antonio Plaza, and Zhihui Wei.
\newblock Anomaly detection in hyperspectral images based on low-rank and
  sparse representation.
\newblock \emph{IEEE Transactions on Geoscience and Remote Sensing},
  54\penalty0 (4):\penalty0 1990--2000, 2015{\natexlab{b}}.

\bibitem[Cheng and Wang(2019)]{cheng2019graph}
Tongkai Cheng and Bin Wang.
\newblock Graph and total variation regularized low-rank representation for
  hyperspectral anomaly detection.
\newblock \emph{IEEE Transactions on Geoscience and Remote Sensing},
  58\penalty0 (1):\penalty0 391--406, 2019.

\bibitem[Li et~al.(2017)Li, Wu, and Du]{liwei2017cnnd}
Wei Li, Guodong Wu, and Qian Du.
\newblock Transferred deep learning for anomaly detection in hyperspectral
  imagery.
\newblock \emph{IEEE Geoscience and Remote Sensing Letters}, 14\penalty0
  (5):\penalty0 597--601, 2017.
\newblock \doi{10.1109/LGRS.2017.2657818}.

\bibitem[Xie et~al.(2019)Xie, Lei, Liu, Li, and Jia]{xie2019spectral}
Weiying Xie, Jie Lei, Baozhu Liu, Yunsong Li, and Xiuping Jia.
\newblock Spectral constraint adversarial autoencoders approach to feature
  representation in hyperspectral anomaly detection.
\newblock \emph{Neural Networks}, 119:\penalty0 222--234, 2019.

\bibitem[Zhao and Zhang(2018)]{zhao2018spectral}
Chunhui Zhao and Lili Zhang.
\newblock Spectral-spatial stacked autoencoders based on low-rank and sparse
  matrix decomposition for hyperspectral anomaly detection.
\newblock \emph{Infrared Physics \& Technology}, 92:\penalty0 166--176, 2018.

\bibitem[Lu et~al.(2019)Lu, Zhang, and Huang]{lu2019exploiting}
Xiaoqiang Lu, Wuxia Zhang, and Ju~Huang.
\newblock Exploiting embedding manifold of autoencoders for hyperspectral
  anomaly detection.
\newblock \emph{IEEE Transactions on Geoscience and Remote Sensing},
  58\penalty0 (3):\penalty0 1527--1537, 2019.

\bibitem[Wang et~al.(2022)Wang, Wang, Zhang, and Zhong]{wangshaoyu2022AutoAD}
Shaoyu Wang, Xinyu Wang, Liangpei Zhang, and Yanfei Zhong.
\newblock Auto-ad: Autonomous hyperspectral anomaly detection network based on
  fully convolutional autoencoder.
\newblock \emph{IEEE Transactions on Geoscience and Remote Sensing},
  60:\penalty0 1--14, 2022.
\newblock \doi{10.1109/TGRS.2021.3057721}.

\bibitem[Zhao et~al.(2017)Zhao, Li, and Zhu]{zhao2017hyperspectral}
Chunhui Zhao, Xueyuan Li, and Haifeng Zhu.
\newblock Hyperspectral anomaly detection based on stacked denoising
  autoencoders.
\newblock \emph{Journal of Applied Remote Sensing}, 11\penalty0 (4):\penalty0
  042605, 2017.

\bibitem[Imani(2017)]{imani2017rx}
Maryam Imani.
\newblock Rx anomaly detector with rectified background.
\newblock \emph{IEEE Geoscience and Remote Sensing Letters}, 14\penalty0
  (8):\penalty0 1313--1317, 2017.

\bibitem[Lei et~al.(2019)Lei, Xie, Yang, Li, and Chang]{lei2019spectral}
Jie Lei, Weiying Xie, Jian Yang, Yunsong Li, and Chein-I Chang.
\newblock Spectral--spatial feature extraction for hyperspectral anomaly
  detection.
\newblock \emph{IEEE Transactions on Geoscience and Remote Sensing},
  57\penalty0 (10):\penalty0 8131--8143, 2019.

\bibitem[Ju et~al.(2018)Ju, Liu, and Wang]{ju2018hyperspetral}
Huihui Ju, Zhigang Liu, and Yang Wang.
\newblock Hyperspetral anomaly detection incorporating spatial information.
\newblock In \emph{2018 Eighth International Conference on Image Processing
  Theory, Tools and Applications (IPTA)}, pages 1--5. IEEE, 2018.

\bibitem[Kang et~al.(2017)Kang, Zhang, Li, Li, Li, and
  Benediktsson]{kang2017hyperspectral}
Xudong Kang, Xiangping Zhang, Shutao Li, Kenli Li, Jun Li, and Jon~Atli
  Benediktsson.
\newblock Hyperspectral anomaly detection with attribute and edge-preserving
  filters.
\newblock \emph{IEEE Transactions on Geoscience and Remote Sensing},
  55\penalty0 (10):\penalty0 5600--5611, 2017.

\bibitem[Bitar et~al.(2019)Bitar, Cheong, and Ovarlez]{bitar2019sparse}
Ahmad~W Bitar, Loong-Fah Cheong, and Jean-Philippe Ovarlez.
\newblock Sparse and low-rank matrix decomposition for automatic target
  detection in hyperspectral imagery.
\newblock \emph{IEEE Transactions on Geoscience and Remote Sensing},
  57\penalty0 (8):\penalty0 5239--5251, 2019.

\bibitem[Huyan et~al.(2018)Huyan, Zhang, Zhou, and
  Jiao]{huyan2018hyperspectral}
Ning Huyan, Xiangrong Zhang, Huiyu Zhou, and Licheng Jiao.
\newblock Hyperspectral anomaly detection via background and potential anomaly
  dictionaries construction.
\newblock \emph{IEEE Transactions on Geoscience and Remote Sensing},
  57\penalty0 (4):\penalty0 2263--2276, 2018.

\bibitem[Zhang et~al.(2015)Zhang, Du, Zhang, and Wang]{zhang2015low}
Yuxiang Zhang, Bo~Du, Liangpei Zhang, and Shugen Wang.
\newblock A low-rank and sparse matrix decomposition-based mahalanobis distance
  method for hyperspectral anomaly detection.
\newblock \emph{IEEE Transactions on Geoscience and Remote Sensing},
  54\penalty0 (3):\penalty0 1376--1389, 2015.

\bibitem[Hanley and McNeil(1982)]{hanley1982meaning}
James~A Hanley and Barbara~J McNeil.
\newblock The meaning and use of the area under a receiver operating
  characteristic (roc) curve.
\newblock \emph{Radiology}, 143\penalty0 (1):\penalty0 29--36, 1982.

\end{thebibliography}






\end{document}